\documentclass[11pt]{article}
\usepackage[english]{babel}    
\usepackage[ansinew]{inputenc} 
\usepackage[T1]{fontenc}       
\usepackage{lmodern,microtype} 
\usepackage{amsmath,amsthm,amssymb,amsfonts} 
\usepackage{dsfont,mathrsfs,accents,bm} 
\usepackage{titlesec,titling}  
\usepackage[nohead]{geometry}  
\usepackage{setspace,caption}  
\usepackage{enumitem,booktabs} 
\usepackage[comma]{natbib} 
\usepackage{pstricks,pst-all}   
\usepackage{hyperref,breakurl}
\usepackage{graphicx}
\usepackage[flushleft]{threeparttable}
\setenumerate{label=\small(\roman*)}
\hypersetup{breaklinks=true,colorlinks=true,pdfnewwindow=true,pdfstartview=FitH,%
	pdftitle="Bounded Covariates",pdfauthor="Kashaev",%
	urlcolor=blue!95!red!50!black,citecolor=blue!95!red!50!black,linkcolor=red!90!black}
\DeclareCaptionFont{fancy}{\bfseries\sffamily} 
\allowdisplaybreaks

\providecommand{\psreset}{\psset{%
		linewidth=0.3pt,linestyle=solid,linecolor=black,
		dotsize=2.5pt,dotsep=2.5pt,arrowsize=4pt,
		fillstyle=none,fillcolor=white,
		showpoints=false,arrows=-,linearc=0,framearc=0,
		hatchsep=2pt,hatchwidth=0.2pt,nodesep=4pt,opacity=1}
	\psset{gridcolor=black!60, subgridcolor=black!30}
}

\psreset
\titleformat{\section}[block]{\centering\large\bfseries\sffamily}{\thesection.}{0.5em}{}
\titleformat{\subsection}[block]{\flushleft\bfseries}{\thesubsection.}{0.5em}{}
\titleformat{\subsection}[block]{\flushleft\bfseries\sffamily}{\thesubsection.}{0.5em}{}
\titleformat{\subsubsection}[runin]{\normalsize\itshape}{\bfseries\upshape\sffamily\thesubsubsection.}{0.5em}{}[.--\:]
\renewcommand{\thesubsubsection}{\arabic{section}.\arabic{subsection}.\alph{subsubsection}}
\titlespacing{\section}{0ex}{10ex}{5ex}
\titlespacing{\subsection}{0in}{6ex}{3ex}
\titlespacing{\subsubsection}{0mm}{2ex}{0.5em}
\pretitle{\vspace*{-10ex}\begin{center}\LARGE\bfseries\sffamily}
	\posttitle{\par\end{center}\vskip 0.5em}
\preauthor{\begin{center} \large \lineskip 0.5em\begin{tabular}[t]{c}}
		\postauthor{\end{tabular}\par\end{center}}
\predate{\begin{center}\small}
	\postdate{\par\end{center} \vspace*{-5ex} }
\providecommand{\abstitle}[1]{{\par\vspace*{2ex}\small\bfseries\sffamily #1}\hspace*{1ex}}
\renewenvironment{abstract}%
{\begin{center}\begin{minipage}{0.82\linewidth}%
			\setlength{\parindent}{0.0em}\abstitle{Abstract}\small}%
		{\end{minipage}\end{center}\vfill\clearpage}
\newtheoremstyle{defn}{3ex}{3ex}{}{}{\bfseries\sffamily}{}{.5em}%
{{\thmname{#1}\:\thmnumber{#2}}\:\thmnote{\mdseries({\small#3})\:}}%
\theoremstyle{defn}
\newtheorem{definition}{Definition}

\newtheorem{assumption}{Assumption}

\newtheoremstyle{prop}{3ex}{3ex}{\itshape}{}{\sffamily\bfseries}{}{.5em}%
{{\thmname{#1}\:\thmnumber{#2}}\:\thmnote{\mdseries({\small#3})\:}}
\theoremstyle{prop}
\newtheorem{proposition}{Proposition}[section]

\newtheorem{corollary}[proposition]{Corollary}
\newtheoremstyle{obs}{2ex}{2ex}{}{}{\itshape}{}{.5em}%
{{\thmname{#1}\:\thmnumber{#2}}\:\thmnote{[{\small#3}]\:}}
\theoremstyle{obs}

\providecommand{\Exp}[1]{\mathds{E}\left[\,#1\,\right]}

\providecommand{\Char}[1]{\mathds{1}\left(\,#1\,\right)}
\providecommand{\Real}{{\mathds{R}}}

\providecommand{\tr}{^{{\sf T}}}
\providecommand{\as}{\ensuremath{\mathrm{a.s.}}}
\providecommand{\rand}[1]{\mathbf{#1}}
\providecommand{\rands}[1]{\boldsymbol{#1}}
\providecommand{\norm}[1]{\left\lVert#1\right\rVert}
\newcommand{\abs}[1]{\left\lvert#1\right\rvert}

\newcommand*{\Scale}[2][4]{\scalebox{#1}{$#2$}}
	\newcommand{\x}{x}
	\newcommand{\y}{y}\newcommand{\Y}{Y}
	\newcommand{\z}{z}
	\newcommand{\w}{w}

	\newcommand{\dop}{h}

\begin{document}

\title{Identification and estimation of multinomial choice models with latent special covariates\thanks{A previous version of this paper was circulated under the title ``Identification and estimation of discrete outcome models with latent special covariates.'' I thank the editor and two anonymous referees for comments and suggestions that have greatly improved the manuscript. I also thank Roy Allen, Victor H. Aguiar, Tim Conley, Nirav Mehta, Salvador Navarro, Joris Pinkse, David Rivers, and Bruno Salcedo for their useful comments and discussions.}}
\author{ 
	Nail Kashaev
	\thanks{Department of Economics, University of Western Ontario.}\\nkashaev@uwo.ca
	}
\date{First version: November 13, 2018\\
 This version: March 21, 2022}

\maketitle

\begin{abstract}
Identification of multinomial choice models is often established by using special covariates that have full support. This paper shows how these identification results can be extended to a large class of multinomial choice models when all covariates are bounded. I also provide a new $\sqrt{n}$-consistent asymptotically normal estimator of the finite-dimensional parameters of the model.  \bigskip  

\bigskip

\end{abstract}

\section{Introduction}
This paper studies identification and estimation of random coefficients multinomial choice models with covariates that have bounded support. Often \emph{some} latent variables in these models have full support (i.e. supported on the whole Euclidean space). Under common restrictions on the distribution of these unobservables, I constructively identify it and show how these latent variables can be used to construct special covariates (i.e., artificial observables with full support) to nonparametrically identify the distribution of \emph{all} the other unobservables. Identification of all parts of the structural model is crucial for welfare analysis (e.g., aggregate welfare changes between two choice situations).
My identification technique is constructive and leads to an asymptotically normal estimator of the finite-dimensional parameters of the model. 
\par
The results of this paper rest on two commonly used assumptions. First, I assume existence of excluded covariates that affect the distribution over choices via a random coefficient. Using variation in these excluded covariates I can identify the distribution of the random coefficient. Second, I assume that the distribution of the random coefficient is sufficiently ``rich''. ``Richness'' of the random coefficient distribution is formalized by a notion of bounded completeness.\footnote{Completeness of a family of distributions is a well-known concept in the Statistics and Econometrics literature. See, for example, \citet{mattner1993some, newey03, chernozhukov2005iv, blundell07, chernozhukov2007instrumental, hu2008instrumental, andrews11, darolles11}, and \citet{d2011completeness}.}
As a result, I show how to identify the distribution over outcomes conditional on the realization of the observed covariates and the latent random coefficient nonparametrically. Since the latent random coefficient often has full support, I can treat it as an observed covariate with full support and apply \emph{any} identification technique that requires existence of such covariates to identify the rest of the model parameters (e.g., the distribution of other latent variables). 
\par
I provide two nonnested identification results. The first result does not make any parametric assumptions about the distribution of latent variables. It, however, imposes some restrictions on the support of observables. In particular, I require the support of some covariates to contain zero. It also requires some smoothness of the distribution of the latent variables. To the best of my knowledge, this is the first result in the literature that nonparametrically identifies the distribution of all latent variables in multinomial choice settings with bounded covariates. The second result uses one of the most popular parameterizations in applied work - a Gaussian distribution of the latent random coefficient. But, in contrast to the first result, it does not require zero in the support of covariates and leaves the distribution of other latent variables completely unrestricted. The second result also leads to an easy to implement asymptotically normal estimator of the finite-dimensional parameters of the model. Similar to \citet{powell1989semiparametric}, this estimator is $\sqrt{n}$-consistent since it is based on average derivatives of an estimable object. 
\par
I contribute to the discrete outcome literature in several respects. I show how existing results that use full-support-excluded covariates with monotonicity restrictions\footnote{See, for example, \citet{manski1985semiparametric, manski1988identification, heckman1990varieties, matzkin1992nonparametric, ichimura1998maximum, lewbel1998semiparametric, lewbel2000semiparametric, tamer03, matzkin2007heterogeneous, berry2009nonparametric, BHR, gautier2013nonparametric, gautier2015triangular, fox2016nonparametric, dunker2017nonparametric, fox2017note, fox2018unobserved, fox2020note}, and \citet{kashaev2020discerning}.} can be directly used in environments with bounded covariates. Formally, I demonstrate that my setting inherits all identifying properties of the setting with a special covariate. I also contribute to the literature on semiparametric models by showing that common parametric restrictions can be used instead of covariates that have full support (e.g., \citealp{fox2012random}). This paper is also related to the literature on identification of finite-dimensional parameters in discrete outcome models with bounded covariates.\footnote{E.g., \citet{magnac2007identification, chen2016informational, kline16}, and \citet{lewbel2021semiparametric}.} The main difference from that literature is that in my framework the distribution of latent variables (e.g., the random intercept) can be nonparametrically identified even if these latent variables have full support, but covariates are bounded. 
\par
My approach is complementary to existing methods. Since as an input my framework requires the average structural demand function (i.e., the choice probability function) for one good, my results may be combined with the ones in \citet{berry2020nonparametric} to nonparametricaly identify the distribution of unobserved individual level heterogeneity. Moreover, in situations where the researcher is not sure whether covariates have full support and is willing to impose mild restrictions because of tractability or data limitations, my approach can provide an additional reassurance of identification. Also, the results in this paper provide a more solid econometric foundation to the models with at least one normally distributed random coefficient (e.g. \citealp{nevo2000practitioner}).
\par
The paper is organized as follows.
In Section~\ref{sec: model}, I describe the setting. Sections~\ref{sec:nonparam} and~\ref{sec:normal error} provide two identification results. 
I show how my identification results can be extended to bundles model in Section~\ref{sub: bundles}. In Sections~\ref{sec: estimation} and~\ref{sec: simulations}, I propose a new estimator of the finite-dimensional parameters and evaluate its performance in simulations. Section~\ref{sec: empirical application} provides an empirical illustration. Section~\ref{sec:conclusion} concludes. All proofs can be found in Appendix~\ref{app: proofs}. Appendix~\ref{app: simulations} provides additional simulation evidence.

\section{Multinomial Choice}\label{sec: model}
Consider the following random coefficients model. The agent maximizes (indirect) utility by choosing between $J$ inside goods (e.g., different brands of cereals) and an outside option of no purchase. The choice set is denoted by $Y=\{0,1,\dots, J\}$. I normalize the utility from alternative $y=0$ to $0$. The random utility from choosing an alternative $y\neq 0$ is\footnote{Deterministic vectors are denoted by lower-case regular font Latin letters (e.g., $x$) and random objects by bold letters (e.g., $\rand{x}$). Capital letters are usually used to denote supports of random variables (e.g., $\rand{x}\in X$). I denote the support of a conditional distribution of $\rand{x}$ conditional on $\rand{z}=z$ by $X_z$.  
The cumulative distribution function (c.d.f.) and the probability density function (p.d.f.) of $\rand{x}$ are denoted by $F_{\rand{x}}$ and $f_{\rand{x}}$. $F_{\rand{x}|\rand{z}}$ ($f_{\rand{x}|\rand{z}}$) denotes the c.d.f. (p.d.f.) of $\rand{x}$ conditional on $\rand{z}=z$. 
}
\begin{align*}\label{eq:random coeff parametrization}
    &\rand{z}_{y}\big[\beta_0(\rand{w})+\beta_{1}(\rand{w})\rand{d}+\rand{e}\big]+\rands{\varepsilon}_y,
\end{align*}
where $\rand{z}_y\in Z_y\subseteq\Real$ is a product-specific observed covariate that can be different for different consumers (e.g., fiber content or price); $\rand{d}\in D\subseteq\Real$ is observed (demographic) individual-specific taste shifter (e.g., age or income); $\rand{w}\in W\subseteq\Real^{d_w}$ is a vector of all other observable covariates, which may include the rest of product/agent characteristics; $\rand{e}\in E\subseteq\Real$ is a latent taste shock. The latent random vector $\rands{\varepsilon}=(\rands{\varepsilon}_{y})_{y\in Y\setminus\{0\}}$ captures all other sources of unobserved heterogeneity (e.g., $\rands{\varepsilon}_y=\rands{\theta}\tr\rand{w}_y+\rands{\epsilon}_y\:\as$, where $\rands{\theta}$ and $\rands{\epsilon}_y$ are random coefficients). The observed covariates are $\rand{x}=(\rand{d}, \rand{z},\rand{\w})$, where $\rand{z}=(\rand{z}_{y})_{y\in Y\setminus\{0\}}$.
\par
The random coefficient $\beta_0(\rand{w})+\beta_{1}(\rand{w})\rand{d}+\rand{e}$ represents individual specific heterogeneous tastes associated with the product characteristic $\rand{z}_{y}$ (i.e., the marginal utility from the product characteristic $\rand{z}_{y}$). This specification of random coefficients is common in applied work (see, for instance, \citealp{berry1995automobile, nevo2000practitioner,nevo2001measuring, berry2004differentiated}). The functions $\beta_0,\beta_1:W\to\Real$ are unknown to the researcher and $\beta_1(w)\neq 0$ for all $w\in W$. I assume that $\rand{d}$ (and $\beta_1(\rand{w})$) is scalar without loss of generality since if $\rand{d}$ is a vector, then all components of it but one can be absorbed by $w$. In this case, one would need to use variation in those absorbed components to identify the coefficients in front of them. Similarly to the existing treatment of random coefficients model, I assume that the random coefficients in front of $\rand{z}_{y}$ are the same for each alternative $y$. However, I do not impose sign restrictions on $\big[\beta_0(\rand{w})+\beta_{1}(\rand{w})\rand{d}+\rand{e}\big]$.\footnote{ Since
$\Pr(\beta_0(\rand{w})+\beta_{1}(\rand{w})\rand{d}+\rand{e}> 0|\rand{x}=x)=1-F_{\rand{e}|\rand{x}}(-\beta_0(w)-\beta_{1}(w)d|x)$ and there are no restrictions on $\beta_0(\cdot)$, the random coefficient $\big[\beta_0(\rand{w})+\beta_{1}(\rand{w})\rand{d}+\rand{e}\big]$ can be positive (negative) with probability that is arbitrarily close to $1$ if the support of $\rand{e}$ conditional on $\rand{x}=x$ is unbounded.}
\par
I start by stating two assumptions that will be used throughout the paper. The first one is a data requirement, the second one is a shape constraint on the distribution of latent variables.

\begin{assumption}[Data]\label{ass:observables}
The analyst can identify $p_0(x)=\Pr(\rand{y}=0|\rand{x}=x)$
for all $x\in X$
\end{assumption}

Assumption~\ref{ass:observables} implies that I only need to observe whether a consumer bought a product or not without knowing the identity of the product (see also, for instance, \citealp{thompson1989identification,lewbel2000semiparametric,fox2012random}).\footnote{The outcome $y=0$ can be replaced by any outcome. In this case, one will just need to renormalize the utility from that outcome to zero.} If the information on the identity of the purchases is also available, then this information (i) may improve the efficiency of an estimator; (ii) can help to satisfy the assumptions needed for identification (e.g., in my empirical illustration, I use one product to identify the sign of $\beta_1$ and I use another one to estimate it); and (iii) can be used to weaken the assumption that the random slope coefficient $\beta_0(\rand{w})+\beta_{1}(\rand{w})\rand{d}+\rand{e}$ is the same across inside goods.

\begin{assumption}[Exclusion Restrictions]\label{ass:exclusion restriction}
For all $w\in W$
\begin{enumerate}
    \item $\rands{\varepsilon}$ is conditionally independent of $(\rand{e},\rand{d},\rand{z})$ conditional on $\rand{w}=w$;
    \item $\rand{e}$ is conditionally independent of $(\rand{d},\rand{z})$ conditional on $\rand{w}=w$.
\end{enumerate}
\end{assumption}
Assumption~\ref{ass:exclusion restriction} is an exclusion restriction that requires latent shocks $\rand{e}$ and $\rands{\varepsilon}$ to be independent of each other (condition (i)) and independent of excluded covariates $(\rand{d},\rand{z})$ (condition (ii)) after conditioning on $\rand{w}$. Assumption~\ref{ass:exclusion restriction} allows any form of dependence between $(\rands{\varepsilon},\rand{e})$ and nonexcluded covariates $\rand{w}$. That is, $\rands{\varepsilon}$ may contain latent product characteristics (e.g., unobserved quality) that can be correlated with nonexcluded covariates (e.g., market-product identifier).\footnote{Since, for identification and estimation, I require the average structural function $p_0$, some forms of endogeneity (i.e., correlation between $\rand{x}$ and $\rands{\varepsilon}$) can be addressed using suitable instruments and control function residuals as in \citet{blundell2004endogeneity} (see also \citealp{berry1994estimating, berry1995automobile, berry2014identification} for identification of structural demand function using aggregate data and instruments).  I leave the detailed analysis of this case for future research.}  In general, since I only require the identification of the structural demand function $p_0$, one can use the results in \citet{berry2020nonparametric} to identify $p_0$ and treat market-product level unobservables as a part of $\rand{w}$.
\par
Next, I provide two nonnested sets of conditions that allow for identification of $\beta_0$, $\beta_1$, and the distribution of $\rand{e}$ and $\rands{\varepsilon}$. In Section~\ref{sec:nonparam}, I impose no parametric assumptions on latent $\rand{e}$ and $\rands{\varepsilon}$ but assume some smoothness on the c.d.f. of $\rands{\varepsilon}$ and restrict the support of covariates. In Section~\ref{sec:normal error}, I identify the model when $\rand{e}$ is normally distributed, without any additional restrictions on the distribution of $\rands{\varepsilon}$ and with minimal support restrictions on covariates. 

\section{Identification}\label{sec: identification}
\subsection{Nonparametric Identification}\label{sec:nonparam}
\begin{assumption}\label{ass:random coeff nonparametric}
For all $w\in W$
\begin{enumerate}
\item Conditional on $\rand{w}=w$, $\rand{e}$ has mean zero and variance one;
\item $F_{\rands{\varepsilon}|\rand{w}}(\cdot|w)$ has bounded partial derivatives up to order $\kappa$ for some $\bar{y}$ and $\partial^l_{\varepsilon^l_{\bar{y}}}F_{\rands{\varepsilon}|\rand{w}}(\cdot|w)|_{\varepsilon=0}\neq 0$ for all $l\leq\kappa$; 
\item There exists $d^*$ such that the support of $(\rand{d},\rand{z})$ conditional on $\rand{w}=w$ contains $(d^*,0)$ with an open neighborhood.
\end{enumerate}
\end{assumption}
Assumption~\ref{ass:random coeff nonparametric}(i) is a scale and location normalization. It restricts $\rand{e}$ conditional on $\rand{w}=w$ to have a finite expectation and a nonzero variance for all $w$. Assumption~\ref{ass:random coeff nonparametric}(ii) requires the conditional  distribution $F_{\rands{\varepsilon}|\rand{w}}$ to be sufficiently smooth in one component of $\varepsilon$ in the neighborhood of zero and have different from zero higher order partial derivatives. Since $\Exp{\rands{\varepsilon}}$ is not assumed to be zero, if, for instance, $\rands{\varepsilon}$ is multivariate normal with component-wise nonzero mean, then Assumption~\ref{ass:random coeff nonparametric}(ii) is automatically satisfied. It is also generically satisfied when at least one component of $\rands{\varepsilon}$ is independent of the others and has a type I extreme value distribution \citep{fox2012random}. However, Assumption~\ref{ass:random coeff nonparametric}(ii) rules out cases when $\rands{\varepsilon}$ is a constant. Another example of violation of Assumption~\ref{ass:random coeff nonparametric}(ii) is when $\kappa$ is infinite and $F_{\rands{\varepsilon}|\rand{w}}$ is a polynomial function of any finite degree. (In Section~\ref{sec:normal error}, I provide an alternative result that does not restrict $F_{\rands{\varepsilon}|\rand{w}}$.) Assumption~\ref{ass:random coeff nonparametric}(iii) requires the support of $\rand{z}$ to contain zero with some open neighborhood. Assumptions similar to Assumptions~\ref{ass:random coeff nonparametric}(ii)-(iii) are common in the literature on identification of random coefficients models (e.g., Assumptions 8 and 10 in \citealp{fox2012random} and Assumption 4 in \citealp{allen2020identification}).

\begin{proposition}\label{prop:nonparameric identification1}
If Assumptions~\ref{ass:observables}-~\ref{ass:random coeff nonparametric} hold, then
$\beta_0(w)$, $\beta_1(w)$, and $\Exp{\rand{e}^l|\rand{w}=w}$, $0\leq l\leq\kappa$, are identified for all $w\in W$.
\end{proposition}
Identification of $\kappa\leq\infty$ moments of the conditional distribution of $\rand{e}$ conditional on $\rand{w}$ is often sufficient for nonparametric identification of it. For example, Assumption 7 in \citet{fox2012random} uses the Carleman condition.\footnote{For more detailed discussion of the problem of identification of the distribution from its moments see, for instance, \citet{kleiber2013multivariate} and references therein.} Thus, under minimal restrictions, I can nonparametically identify the conditional c.d.f. $F_{\rand{v}|\rand{x}}$, where $\rand{v}=\beta_0(\rand{w})+\beta_{1}(\rand{w})\rand{d}+\rand{e}$. 
\par
To establish the next identification result I need the following definition.
\begin{definition}[Bounded completeness]\label{def:bounded completeness general}
The family of distributions $\left\{F_{\rand{v}|\rand{x}}(\cdot|x),x\in X'\right\}$ is boundedly complete if
\[
\forall x\in X',\:\int_{V}g(t)dF_{\rand{v}|\rand{x}}(t|x)=0\implies g(\rand{v})=0 \:\as,
\]
for any bounded function $g$.
\end{definition}
Completeness assumptions have been widely used in econometric analysis. Completeness is typically imposed on the distribution of observables (e.g., \citealp{newey03}). However, many commonly used parametric restrictions on the distribution of unobservables imply bounded completeness. For instance, it is satisfied for normal distributions and the Gumbel distribution.\footnote{For testability of the completeness assumptions see \citet{canay13}.}
\par
Combining bounded completeness with the identified distribution of the index $\rand{v}$, I have the following result.
\begin{proposition}\label{prop:nonparameric identification2}
If $F_{\rand{v}|\rand{x}}$ is identified and
    $
    \left\{F_{\rand{v}|\rand{x}}(\cdot|(d,z,w)),d\in D_{(z,w)}\right\}
    $
    is boundedly complete for all $(z,w)$ in the support, then the above model inherits all identifying properties of the random coefficients model with utilities 
    $
        \Char{y\neq0}(\rand{r}_{y}+\rands{\varepsilon}_{y}).
    $
    The vector $\rand{r}=(\rand{r}_{y})_{y\in Y\setminus\{0\}}$ is an observed covariate conditionally independent of $\rands{\varepsilon}=(\rands{\varepsilon}_y)_{y\in Y\setminus\{0\}}$ conditional on $\rand{w}=w$ with the conditional support 
    $
    R_w=\left\{r\in\Real^J\::\:r=v z,\: z\in Z_w, v\in V_w\right\},
    $
    where $V_w$ is the support of $\rand{v}$ conditional on $\rand{w}=w$.
    In particular, $F_{\rands{\varepsilon}|\rand{w}}$ is identified over $R_w$.
\end{proposition}
The proof of Proposition~\ref{prop:nonparameric identification2} is similar to the proof of Theorem 11 in \citet{fox2012random}. The main difference is that, instead of parametric restrictions, Proposition~\ref{prop:nonparameric identification2} uses the interaction between $\rand{d}$ and $\rand{z}$.
\par
Proposition~\ref{prop:nonparameric identification2} implies that the original random coefficient model can be represented in the ``special-covariate-with-full-support'' framework without assuming existence of such covariates. Moreover, if the set of directions that $z/\norm{z}$ can cover is sufficiently rich and the support of $\rand{e}$ conditional on $\rand{w}=w$ is $\Real$, then $R_{w}=\Real^{J}$ and all the identification results that require existence of special covariates with full support (e.g., \citealp{lewbel2000semiparametric, berry2009nonparametric, gautier2015triangular, fox2016nonparametric}, and \citealp{fox2020note}) can be applied. 
\par
Combining the results in Propositions~\ref{prop:nonparameric identification1} and~\ref{prop:nonparameric identification2} with Theorem~1 in \citet{fox2020note}, I can establish the following result.
\begin{corollary}\label{corr: foxnote}
For all $y\neq 0$, let $\rands{\varepsilon}_y=\rands{\theta}\tr\rand{w}_y+\rands{\zeta}_y$, where $\rands{\theta}$ and $\rands{\zeta}=(\rands{\zeta}_y)_{\y\in\Y\setminus\{0\}}$ are random coefficients, and $\rand{w}_y$ is the vector of product-$y$-specific covariates. Suppose 
\begin{enumerate}
\item The assumptions of Propositions~\ref{prop:nonparameric identification1} and~\ref{prop:nonparameric identification2} hold; 
\item $R_{\w}=\Real^{J}$ for all $w\in W$;
\item $(\rands{\theta},\rands{\zeta})$ and $\rand{w}=(\rand{w}_y)_{\y\in\Y\setminus\{0\}}$ are independent;
\item The support of $\rand{w}$ contains an open ball of dimensionality of $\rand{w}$; 
\item $(\rands{\theta},\rands{\zeta})$ has finite absolute moments and its distribution is uniquely determined by its moments;
\end{enumerate}
then $\beta_0$, $\beta_1$, and the distribution of $(\rand{e}_1,\rands{\theta},\rands{\epsilon})$ are identified.
\end{corollary}

To the best of my knowledge, Corollary~\ref{corr: foxnote} is the first result that establishes nonparametric identification of the whole distribution of the random coefficients in the multinomial choice environments without assuming the existence of special covariates. \citet{fox2012random,allen2020identification}, and \citet{lewbel2021semiparametric} also allow for bounded covariates. However, they either do not fully identify the distribution of the random intercept $\rands{\varepsilon}$ \citep{allen2020identification,lewbel2021semiparametric} or impose parametric restrictions on it \citep{fox2012random}.

\subsection{Normal Taste Shock}\label{sec:normal error}
\begin{assumption}\label{ass:normal error random coef}
For all $w\in W$
\begin{enumerate}
\item Conditional on $\rand{w}=w$, $\rand{e}$ is a standard normal random variable;
\item there exists $(d^*,z^{*\sf T})$ in the interior of the support of $(\rand{d},\rand{z})$ conditional on $\rand{w}=w$ such that $\z^*_{y}>0$ for all $y\in Y$;
\item there exists $(d^{**},z^{**\sf T})$ in the interior of the support of $(\rand{d},\rand{z})$ conditional on $\rand{w}=w$ such that $p_0((d,z^{**},w))$ is neither an exponential nor an affine function of $d$ on some open set.
\end{enumerate}
\end{assumption}
Assumption~\ref{ass:normal error random coef}(i) requires $\rand{e}$ to be normally distributed with nonzero variance. With nonzero variance, the assumption that $\Exp{\rand{e}^2}=1$ is just a scale normalization. The assumption is common in applied work (e.g., \citealp{nevo2000practitioner,nevo2001measuring}) and allows me to relax Assumptions~\ref{ass:random coeff nonparametric}(ii)-(iii). Assumption~\ref{ass:normal error random coef}(ii) is only needed for identification of the sign of $\beta_1(w)$. Assumption~\ref{ass:normal error random coef}(iii) means that if I fix all covariates but the one that shifts the random coefficient, then the probability of the default conditional on covariates is neither an affine nor an exponential function of this nonfixed covariate. Assumption~\ref{ass:normal error random coef}(iii) is not very restrictive since it rules out only some exponential and linear probability models. Moreover, it is testable. 

\begin{proposition}\label{prop: normal mult choice}
If Assumptions~\ref{ass:observables},~\ref{ass:exclusion restriction}, and~\ref{ass:normal error random coef} hold, then
\begin{enumerate}
    \item $\beta_0(w)$ and $\beta_1(w)$ are identified for all $w\in W$;
    \item The conditions of Proposition~\ref{prop:nonparameric identification2} are satisfied.
\end{enumerate}
\end{proposition}
The proof of the identification of $\beta_{0}$ and $\beta_{1}$ uses the multiplicative structure of $d$ and $z$, and properties of the standard normal p.d.f. Informally, note that
\[
\beta_{0}(\rand{w})\rand{z}+\beta_{1}(\rand{w})\rand{d}\rand{z}+\rand{e}\rand{z}.
\]
Since $d$ and $z$ can be moved independently, I can use variation in $d$ while keeping $dz$ by varying $z$ to identify $\beta_{0}(w)$. Then, by varying $z$, I can identify $\beta_{1}(w)$. Proposition~\ref{prop: normal mult choice}(ii) follows from $\beta_0(w)$ and $\beta_1(w)$ being identified and $\rand{e}$ being standard normal (i.e, $\beta_0(\rand{w})+\beta_1(\rand{w})\rand{d}+\rand{e}$ conditional on $\rand{x}=x$ generates a boundedly complete family of distributions).
\par
Note that the only restriction on $\rands{\varepsilon}$ needed for Proposition~\ref{prop: normal mult choice} is the conditional independence assumption (Assumption~\ref{ass:exclusion restriction}). The random intercept $\rands{\varepsilon}$ is allowed be continuously or discretely distributed (e.g., it may be a constant). Hence, I can extend Theorem 2 in \citet{fox2016nonparametric} to environments with bounded covariates.
\begin{corollary}
For all $y\neq 0$ let $\rands{\varepsilon}_y=\rands{\theta}_y(\rand{w})$, where $\rands{\theta}_{y}$ is a random function such that its realization $\theta_y$ is a map from $W$ to $\Real$. Suppose 
\begin{enumerate}
\item Assumptions of Proposition~\ref{prop: normal mult choice} hold; 
\item $R_{w}=\Real^{J}$ for all $w\in W$;
\item $\rands{\theta}=(\rands{\theta}_y)_{y\neq 0}$ and $\rand{w}$ are independent;
\item The support of $\rands{\theta}$, $\Theta$, satisfies Assumption 4 in \citet{fox2016nonparametric};
\end{enumerate} 
then $\beta_0$, $\beta_1$, and the distribution of $\rands{\theta}$ are identified.
\end{corollary}

\subsection{Bundles}\label{sub: bundles}
Note that since I do not assume independence among $\rands{\varepsilon}_{y}$ across $y$, the multinomial choice model I study covers some bundles models \citep{gentzkow2007valuing, dunker2017nonparametric,fox2017note}. In particular, assume that there are $\tilde{J}$ goods and the agent can purchase any bundle consisting of these goods. The vector $\tilde{y}$ describes the purchasing decision of the agent. That is, $\tilde{y}\in\tilde{Y}=\{0,1\}^{\tilde{J}}$. For instance, $\tilde{y}=(0,1,0,1,0,\dots,0)$ corresponds to the case when the agent purchased a bundle of goods $2$ and $4$. The random utility from choosing bundle $\tilde{y}\neq 0$ is of the form 
\begin{align*}
    (\beta_0(\rand{w})+\beta_{1}(\rand{w})\rand{d}+\rand{e}) \sum_{j=1}^{\tilde{J}} \tilde{y}_j \rand{\tilde{z}}_{j}+\rands{\varepsilon}_{\tilde{y}},
\end{align*}
and the utility from buying nothing is zero. I can rewrite the above utilities from bundles as as the utilities form the multinomial choice problem since there are finitely ($2^{\tilde{J}}$) possible bundles. Indeed, I can enumerate them all with $y=0$ corresponding to $\tilde{y}=0\in\Real^{\tilde{J}}$ (i.e., $Y=\{0,1,2,\dots,2^{\tilde{J}}\}$) and define $z_y=\sum_{j=1}^{\tilde{J}} \tilde{y}_j \tilde{z}_{j}$. As a result, I can extend the conclusions of Theorem 1 in \citet{fox2017note} to environments with bounded covariates
\begin{corollary}
Let $J=2$ and
\begin{align*}
    \rands{\varepsilon}_{(1,0)}&=\theta_1(\rand{w})+\rands{\epsilon}_1,\quad\quad
    \rands{\varepsilon}_{(0,1)}=\theta_2(\rand{w})+\rands{\epsilon}_2,\\
    \rands{\varepsilon}_{(1,1)}&=\rands{\varepsilon}_{(1,0)}+\rands{\varepsilon}_{(0,1)}+\rands{\xi}\theta_3(\rand{w}),
\end{align*}
where $\theta_i(\cdot)$, $i=1,2,3$, are some unknown functions, and $(\rands{\epsilon}_1,\rands{\epsilon}_2,\rands{\xi})\in\Real^2\times\Real_{+}$. Suppose
\begin{enumerate}
    \item Assumptions of Propositions~\ref{prop:nonparameric identification1} and~\ref{prop:nonparameric identification2} or Proposition~\ref{prop: normal mult choice} hold;
    \item $R_w=\Real$ for all $w$;
    \item $(\rands{\epsilon}_1,\rands{\epsilon}_2)|\rand{w}=w$ has an everywhere positive Lebesgue density on its support for all $w\in W$;
    \item $\Exp{\rands{\epsilon}_i|\rand{w}=w}=0$ and $\Exp{\rands{\xi}|\rand{w}=w}=1$ for all $w\in W$ and $i=1,2$,
\end{enumerate}
then $\theta_i(\cdot)$, $i=1,2,3$, and the c.d.fs $F_{\rands{\epsilon}_i|\rand{w}}$, $i=1,2$, and $F_{\rands{\xi}|\rand{w}}$ are identified.
\end{corollary}

\section{Estimation of \texorpdfstring{$\beta$}{the Finite-Dimensional parameter}}\label{sec: estimation}
Proposition~\ref{prop: normal mult choice} constructively identifies $\beta_{0}$ and $\beta_{1}$. In this section, I use it to estimate these parameters. That is, I focus on the multinomial choice model with random coefficients with normally distributed $\rand{e}$.\footnote{Proposition~\ref{prop:nonparameric identification1} also provides a constructive identification for $\beta_0$ and $\beta_1$. However, Assumption~\ref{ass:random coeff nonparametric}(iii) fails to hold in my illustrative application presented in Section~\ref{sec: empirical application}. Additionally, Proposition~\ref{prop:nonparameric identification1} uses limits of derivatives of identifiable functions at a single point, thus, most likely, leading to a consistent estimator with nonparametric rate of convergence.} Moreover, to simplify the exposition, I assume that there are no nonexcluded covariates $\rand{w}$ (i.e., $\beta_0(\cdot)$ and $\beta_1(\cdot)$ are constant functions). Note that, even though $\beta_0$ and $\beta_1$ are finite-dimensional parameters and the distribution of $\rand{e}$ is assumed to be known, the model is still semiparametric since the distribution of $\rands{\varepsilon}$ is not parametric.
\par
The first ingredient of the estimator is a nonparametric estimator of $p_0(\cdot)=\Pr(\rand{y}=0|\rand{x}=\cdot)$, $\hat{p}_0(\cdot)$. Any consistent and smooth enough estimator $\hat{p}_0$ will deliver a consistent estimator of $\beta=(\beta_1,\beta_0)$.\footnote{The normality of $\rand{e}$ implies that $p_0$ has continuous derivatives of any order. See Appendix~\ref{app: proof of est} for details.} For concreteness, I work with the series estimator based on products of powers of components of $x=(d,z)$ (polynomial regressions). That is, given a sample of independent identically distributed (i.i.d.) observations on covariates and a binary random variable that indicates whether the product was purchased or not $\left\{\Char{\rand{y}^{(i)}=0},\rand{x}^{(i)}\right\}_{i=1}^n$, define
\begin{align*}
\hat{p}_0(x)&=\psi^K(x)\tr \left(\Psi\tr\Psi\right)^{-}\sum_{i=1}^n\psi^K\left(\rand{x}^{(i)}\right)\Char{\rand{y}^{(i)}=0},
\end{align*}
where $\psi^K(\cdot)$ is a vector of orthonormal basis functions based on products of powers of components of $x$,
$\Psi=\left(\psi^K\left(\rand{x}^{(1)}\right),\psi^K\left(\rand{x}^{(2)}\right),\dots,\psi^K\left(\rand{x}^{(n)}\right)\right)\tr$, and $\left(\Psi\tr\Psi\right)^{-}$ is the Moore-Penrose generalized inverse. I assume that the sum of powers of components of $x$ in $\psi^K$ is monotonically increasing in $K$.
\par
The sign of $\beta_1$ can be trivially estimated from $\hat{p}_0$ since 
\[
\mathrm{sign}(\beta_1)=\mathrm{sign}\left(p_0((d',z))-p_0((d,z))\right)\mathrm{sign}(z_{y^*})\mathrm{sign}(d'-d)
\]
if $z\geq 0$ or $z\leq 0$ with $z_{y^*}\neq 0$. Hence, for simplicity I assume that $\beta_1>0$.
\par
The identification result in Proposition~\ref{prop: normal mult choice} is constructive and provides a closed form expression for $\beta$ as a functional of $p_0$ (see Appendix~\ref{app: proof of est}). Given the nonparametric power series estimator $\hat{p}_0$, the plug-in estimator of $\beta$ is
\begin{align*}\label{eq:beta^2}
    \hat{\beta}_1&=\sqrt{\frac{\displaystyle\sum\limits_{i=1}^n\hat{p}_{111}\left(\rand{x}^{(i)}\right)\hat{p}_1\left(\rand{x}^{(i)}\right)-\hat{p}_{11}\left(\rand{x}^{(i)}\right)^2}{\displaystyle\sum\limits_{i=1}^n\hat{p}_{12}\left(\rand{x}^{(i)}\right)\hat{p}_1\left(\rand{x}^{(i)}\right)-\hat{p}_2\left(\rand{x}^{(i)}\right)\hat{p}_{11}\left(\rand{x}^{(i)}\right)-\hat{p}_1\left(\rand{x}^{(i)}\right)^2}},\\
    \hat{\beta}_0&=\hat{\beta}_1\dfrac{\displaystyle\sum\limits_{i=1}^n\hat{p}_{2}\left(\rand{x}^{(i)}\right)-\rand{d}^{(i)}\hat{p}_{1}\left(\rand{x}^{(i)}\right)}{\displaystyle\sum\limits_{i=1}^n\hat{p}_{1}\left(\rand{x}^{(i)}\right)}-\frac{1}{\hat{\beta}_1}\dfrac{\displaystyle\sum\limits_{i=1}^n\hat{p}_{11}\left(\rand{x}^{(i)}\right)}{\displaystyle\sum\limits_{i=1}^n\hat{p}_{1}\left(\rand{x}^{(i)}\right)},
\end{align*}
where
\begin{align*}
    \hat{p}_{1}(x)&=\partial_{d}\hat{p}_0(x),\quad
    \hat{p}_{11}(x)=\partial^2_{d^2}\hat{p}_0(x),\quad
    \hat{p}_{111}(x)=\partial^3_{d^3}\hat{p}_0(x),\\
    \hat{p}_2(x)&=\sum_{y=1}^{J}z_{y}\partial_{z_{y}}\hat{p}_0(x),\quad
    \hat{p}_{12}(x)=\partial_{d}\hat{p}_2(x).
\end{align*}
Note that $\hat\beta$ is essentially a nonlinear function of sample averages of different derivatives of estimated $\hat{p}_0$. Following \citet{newey1994asymptotic, newey1997convergence}, to achieve $\sqrt{n}$-consistency and asymptotic normality of the proposed estimator, I will have to establish existence of the Reisz representer of a particular directional derivative. Let
\begin{align*}
    \bar{v}_1(x)&=-\Big[4p_{1111}(x)f_{\rand{x}}(x)+8p_{111}(x)\partial_{d}f_{\rand{x}}(x)+5p_{11}(x)\partial^2_{d^2}f_{\rand{x}}(x)+p_{1}(x)\partial^3_{d^3}f_{\rand{x}}(x)\Big]/f_{\rand{x}}(x),\\
    \bar{v}_2(x)&=\Big[\beta_1\{(1-J)f_{\rand{x}}(x)+d\partial_{d}f_{\rand{x}}(x)-\sum_{y}z_{y}\partial_{z_{y}}f_{\rand{x}}(x)\}-\partial^2_{d^2}f_{\rand{x}}(x)\Big]/f_{\rand{x}}(x),\\
    \bar{v}(x)&=(\bar{v}_1(x),\bar{v}_2(x)),
\end{align*}
where $f_{\rand{x}}$ is the p.d.f. of $\rand{x}$, and $p_1$, $p_{11}$, $p_{111}$, and $p_{1111}$ are first, second, third, and forth derivatives of $p_0$ with respect to $d$, respectively.
\begin{assumption}\label{ass: regularity}
\begin{enumerate}
    \item The support of $\rand{x}$, $X$, is a Cartesian product of compact connected nonsingleton intervals in $\Real$.
    \item $f_{\rand{x}}$ is bounded away from zero on the interior of $X$;
    \item $f_{\rand{x}}$, $\partial_{d}f_{\rand{x}}$, $\partial_{z_{y}}f_{\rand{x}}$, and $\partial^2_{d^2}f_{\rand{x}}$ equal to zero at the boundary of $X$ for all $y$;
    \item $\Exp{\bar{v}(\rand{x})\bar{v}(\rand{x})\tr}$ is finite and nonsingular.
\end{enumerate}
\end{assumption}
Assumptions~\ref{ass: regularity}(i)-(ii) are standard in the literature on nonparametric estimation of conditional expectations.  
Similarly to the average derivative estimator of \citet{powell1989semiparametric}, to achieve $\sqrt{n}$-consistency the estimator I need to impose restrictions on the behavior of $f_{\rand{x}}$ on the boundary of its support. Since \citet{powell1989semiparametric} work with the first derivative they only require $f_{\rand{x}}$ to vanish on the boundary. My estimator involves derivatives up to order 3, thus, leading to Assumption~\ref{ass: regularity}(iii). Assumption~\ref{ass: regularity}(iv) is the mean-square continuity condition that requires the variance of the score function of $\rand{x}$ (i.e $\log f_{\rand{x}}$) and derivatives of it to be finite.
\par
The following proposition establishes asymptotic normality of my estimator and is based on Theorem 6 in \citet{newey1997convergence}. Denote
\[
G=\left(\begin{array}{cc}
    2\beta_1 &  0\\
    0 & 1
\end{array}\right)\left(\begin{array}{cc}
    {\Exp{p_{12}(\rand{x})p_{1}(\rand{x})-p_{2}(\rand{x})p_{11}(\rand{x})-p_{1}(\rand{x})^2}} &  0\\
    0 & {\beta_1\Exp{p_{1}(\rand{x})}}
\end{array}\right)^{-1},
\]

\begin{proposition}\label{prop: estimator}
If (i) $\left\{\Char{\rand{y}^{(i)}=0},\rand{x}^{(i)}\right\}_{i=1}^n$ are i.i.d.; (ii) Assumptions~\ref{ass:exclusion restriction},~\ref{ass:normal error random coef} and~\ref{ass: regularity} are satisfied, and Assumption~\ref{ass:normal error random coef}(iii) is satisfied for all $x^{**}=(d^{**},z^{**})\in X$; (iii) $K^6/n\to_{n\to\infty}0$, then
\[
\sqrt{n}(\hat\beta-\beta)\to_{d} \mathrm{N}(0,V),
\]
where $V=G\Exp{\bar{v}(\rand{x})\bar{v}(\rand{x})\tr p_0(\rand{x})(1-p_0(\rand{x}))}G\tr$.
\end{proposition}
In the proof of Proposition~\ref{prop: estimator}, I also provide a consistent estimator of the asymptotic variance matrix $V$ that is based on the estimator proposed in \citet{newey1997convergence}.
\par
I conclude this section by noting that after $\beta$ is estimated, one can construct a sieve maximum-likelihood estimator of $F_{\rands{\varepsilon}}$ since 
\[
\Pr(\rand{y}=0|\rand{x}=x)=\int_\Real F_{\rands{\varepsilon}}(tz_1,tz_2,\dots,tz_{J})\phi \left(t+\beta_0+\beta_1d\right)dt
\]
where $\phi(\cdot)$ is the standard normal p.d.f. Thus, one can find the maximizer of
\begin{align*}
\max_{F\in\mathcal{F}_n}\sum_{i=1}^n &\Char{\rand{y}^{(i)}=0}\log\left(\int_\Real F(tz_1,tz_2,\dots,tz_{J})\phi \left(t+\hat{\beta}_0+\hat{\beta}_1d\right)dt\right)+\\
&\Char{\rand{y}^{(i)}\neq0}\log\left(1-\int_\Real F(tz_1,tz_2,\dots,tz_{J})\phi \left(t+\hat{\beta}_0+\hat{\beta}_1d\right)dt\right),
\end{align*}
where $\{\mathcal{F}_n\}_{n=1}^{\infty}$ is a sequence of sieve spaces for $F_{\rands{\varepsilon}}$. Inference on known functionals of $\beta$ and $F_{\rands{\varepsilon}}$ (e.g., counterfactuals) can be done using likelihood-ratio type statistic (see, for instance, \citealp{shen2005sieve,chen2014sieve}).\footnote{Both $\beta$ and $F_{\rands{\varepsilon}}$ can be estimated in one step by the sieve maximum-likelihood estimator. In this case, however, the estimator of $\beta$ may not be $\sqrt{n}$ consistent.}
\section{Monte-Carlo Simulations}\label{sec: simulations}
In this section, I assess the performance of my estimator in finite samples. I consider the binary choice model:
\[
\rand{y}=\Char{(\beta_0+\beta_1\rand{d}+\rand{e})\rand{z}+\beta_3+\rands{\varepsilon}\geq 0},
\]
where $\beta_0=-0.5$, $\beta_1=1$, and $\rand{e}$ is a standard normal random variable. The random intercept $\beta_3+\rands{\varepsilon}$ is independent from $\rand{x}$ and $\rand{e}$ with mean $\beta_3=0.5$. The observed covariates $\rand{x}=(\rand{d},\rand{z})$ are distributed according to a monotone transformation of a bivariate normal distribution: $\rand{x}=5(\mathrm{arctan}(\rand{\tilde{x}})/\pi+0.5)$, where $\rand{\tilde{x}}$ is a mean-zero normal random vector such that each component of it has variance 1 and the correlation between components is $0.1$. Note that $\rand{x}$ has bounded support. 
\par
I consider several data generating processes (DGPs). The first one (DGP-$0$) is when $\rands{\varepsilon}$ is a standard normal random variable. The next five DGPs correspond to $\rands{\varepsilon}$ being an equally weighted  mixture of three unit-variance normal distributions with mean $-t$, $0$, and $t$ for $t\in\{1,2,3,4,5\}$ (DGP-$t$). For every $t$ the distribution of $\rands{\varepsilon}$ is symmetric. However, the variance is growing with $t$ and the distribution changes from a unimodal distribution to a distribution with three modes. Finally, DGP-L corresponds to the case with logistically distributed $\rands{\varepsilon}$.
\par
Each experiment is conducted $1000$ times for every DGP for 3 sample sizes $n\in\{10^3,5\cdot10^3,10^4\}$. I use a tensor product of cubic polynomials in estimation of the conditional probability $p_0$.\footnote{The results are qualitatively the same for higher order polynomials.} The results for the mean deviation (bias) of the estimator of $\beta_1$ are presented in Table~\ref{table: bias my}. As expected, the bias decreases with the sample size.\footnote{The mean absolute deviation of the estimator also decreases with the sample size. See, Appendix~\ref{app: simulations} for further details.} However, there is not much variation across DGPs.\footnote{For comparison of my estimator with two alternative potentially misspecified parametric estimators, see Appendix~\ref{app: simulations}.} 
\begin{table}[h]
\centering
\begin{threeparttable}
\centering
\caption{Bias}\label{table: bias my}
\begin{tabular}{c|ccccccc}
\hline
\hline
Sample Size & DGP-$0$ & DGP-$1$ & DGP-$2$ & DGP-$3$ & DGP-$4$ & DGP-$5$ & DGP-L\\ 
\hline
1000            & 1.08    & 1.10    & 1.18    & 1.46    & 1.51    & 1.50    & 1.12 \\
5000            & 0.36    & 0.54    & 0.89    & 1.05    & 1.25    & 1.23    & 0.62 \\
10000           & 0.17    & 0.26    & 0.57    & 0.84    & 1.09    & 1.17    & 0.38 \\
\hline
\end{tabular}
\vspace{1ex}
\end{threeparttable}
\end{table}

\section{Illustrative Empirical Application}\label{sec: empirical application}
To illustrate the empirical importance of the relaxation of the parametric assumptions about the distribution of $F_{\rands{\varepsilon}}$ and the proposed estimation procedure, I analyze margarine purchasing decisions of households from Springfield, MO, USA, using the multinomial choice model with normally distributed $\rand{e}$. I find substantial differences between estimates obtained by employing my semiparametric estimator and a fully parametric multinomial-logit-type estimator. 

\subsection*{Data}
The original dataset, constructed by \citet{allenby1991quality}, is a panel of 9196 purchases of 10 brands of stick and tube margarine by 517 households from Springfield, MO, USA, extracted from an ERIM (A.C. Nielsen) scanner dataset. The dataset contains information on the shelf prices of each brand that is constructed using the actual price paid and the value of any redeemed coupon. The household demographics contain information on the household income.\footnote{See \citet{allenby1991quality} for specific details of the dataset construction.} \citet{benoit2016outlier} focused on 5 brands instead of 10 and transformed this dataset to a cross-section with 242 households. In particular, every observation contains only information on the household annual income, which I use as the agent-specific covariate $d$, agent choices ($y$), and product-specific prices $p_y$.\footnote{Income and prices are measured in thousands of US dollars and US dollars, respectively.} There are 5 brands: Generic ($y=0$), Blue Bonnet ($y=1$), House Brand ($y=2$), Shed Spread ($y=3$), and Fleischmann's ($y=4$).
\par
Income varies from 2.5k to 130k, with the median and average income being 26.75k and 22.5k, respectively. Table~\ref{table: sum1} summarizes  the share and price information for different products. There is a variation in prices across brands with Generic being on average the cheapest and Fleischmann's being the most expensive. At the same time, Fleischmann's is the least demanded product.
\begin{table}[h]
\centering
\begin{threeparttable}
\centering
\caption{Summary Statistics for Products}\label{table: sum1}
\begin{tabular}{c|ccccc}
\hline
\hline
Brand         & Share & Average Price & Median Price & Min Price & Max Price\\ 
\hline
Generic       & 0.17  & 0.37          & 0.36         & 0.33      & 0.53\\
Blue Bonnet   & 0.30  & 0.58          & 0.61         & 0.19      & 0.76\\
House Brand   & 0.19  & 0.51          & 0.57         & 0.19      & 0.58\\
Shed Spread   & 0.21  & 0.83          & 0.85         & 0.50      & 0.98\\
Fleischmann's & 0.12  & 1.04          & 1.08         & 0.99      & 1.13\\
\hline
\end{tabular}
\vspace{1ex}
\end{threeparttable}
\end{table}
\subsection*{Utility}
I follow \citet{nevo2000practitioner,nevo2001measuring} and model the utility from purchasing brand $y\in\{0,1,2,3,4\}$ as 
\[
\rands{\delta}\rand{d}+(\beta_0+\beta_1 \rand{d}+\rand{e}) \rand{p}_{y}+\tilde{\rands{\varepsilon}}_y.
\]
The random coefficient $\rands{\delta}$ captures the direct marginal effect of income on utility from consumption of margarine (i.e., it is the same for all brands). The coefficient $\beta_0+\beta_1 \rand{d}$ can be thought of as the average marginal utility with respect to price. It captures the sensitivity of agents with respect to prices and is expected to be negative. Agents with different incomes may react differently to price changes. Note that no assumptions are made about $\tilde{\rands{\varepsilon}}_y$ (e.g., it is not assumed that it is has zero mean).\footnote{Estimation using $\log(\rand{p}_y)$ instead of $\rand{p}_y$ gives qualitatively similar results.} This utility specification correspond to the ``preference shifter'' specification in \citet{griffith2018income}.
There is no information about those who did not purchase any margarine products, thus, I analyze the choices of those who already decided to purchase a margarine product. 
If I treat the utility from consuming Generic brand as the baseline utility and subtract it from all utilities, the normalized utility from purchasing different brands for $y=1,2,3,4$ is
\[
(\beta_0+\beta_1 \rand{d}+\rand{e}) [\rand{p}_{y}-\rand{p}_{0}]+\tilde{\rands{\varepsilon}}_y-\tilde{\rands{\varepsilon}}_0,
\]
and the utility from purchasing Generic brand is $0$. Hence, I can define $\rand{z}_{y}=\rand{p}_{y}-\rand{p}_{0}$ and $\rands{\varepsilon}_y=\tilde{\rands{\varepsilon}}_y-\tilde{\rands{\varepsilon}}_0$, $y=1,2,3,4$, where $\rand{p}_0$ is the price of Generic margarine. 
\par
Given that I am considering margarine products, it is not surprising that the support for $\rand{z}_y$ is far from being full. In particular, $\max_{y}\max_{i}z^{(i)}_y=0.78$ and  $\min_{y}\min_{i}z^{(i)}_y=-0.15$. At the same time, there is still variation in relative prices $\rand{z}_y$ and income $\rand{d}$. This variation allows me to recover $\beta$ without specifying the distribution of $\rands{\varepsilon}$. 
\par
In the current application, I use a minimal amount of information: there are only two covariates. If one has more demographic and product data, it can be easily incorporated into the current framework via $\rand{w}$. For instance,  $\rand{w}$ may contain nonprice marketing variables, packet size dummies, saturated fat content, household size, age of the household head, household location (e.g. zip-code).

\subsection*{Parametric Estimation}
First, I assume the most common parametric specification for the random intercept -- multinomial logit. Formally, I estimate the following specification for normalized utility:
\[
\Char{y\neq0}[\gamma_y+(\beta_0+\beta_1 \rand{d}+\rand{e}) \rand{z}_{y}]+\alpha\rands{\varepsilon}_y,
\]
where $\{\rands{\varepsilon}_y\}_{y=0}^4$ are i.i.d. Gumbel across $y$ that are also independent from $\rand{x}=(\rand{d},\rand{z})$; $\rand{e}$ is a standard normal random variable. (Parameter $\alpha$ captures the scale of $\rands{\varepsilon}_y$ since the variance of $\rand{e}$ is set to 1.) Although, price $\rand{p}_y$ is probably correlated with unobserved part of the utility $\tilde{\rands{\varepsilon}}_y$ (e.g., unobserved quality), the price difference $\rand{z}_y=\rand{p}_y-\rand{p}_0$ may be independent from $\tilde{\rands{\varepsilon}}_y-\tilde{\rands{\varepsilon}}_0$.
\par
The estimates of $\beta_0$ and $\beta_1$ are $\bar{\beta}_0=-6331.94$ (standard error$=17.19$) and $\bar{\beta}_1=-19.69$ (standard error$=514.48$), respectively. As expected, the sign of $\bar{\beta}_0$ is negative. The coefficient in front of the income variable, $\bar{\beta}_1$, is negative and not significant at the $5$ percent significance level. Although income does not matter much, the overall sensitivity to prices (mostly captured by $\bar\beta_0$ in this case) is substantial. The effect of income on marginal disutility from the price increase is not surprising given that margarine constitutes a small share of household expenditures on groceries.\footnote{E.g., in UK households spend about one percent of their grocery expenditures on margarine and butter \citep{griffith2018income}.}

\subsection*{Semiparametric Estimation}
Next, I apply the estimator proposed in Section~\ref{sec: estimation}. Formally, I estimate the following specification for normalized utility:
\[
\Char{y\neq0}[(\beta_0+\beta_1 \rand{d}+\rand{e}) \rand{z}_{y}+\rands{\varepsilon}_y],
\]
where $\rand{e}$ is a standard normal random variable. The random intercept $\rands{\varepsilon}=(\rands{\varepsilon}_y)_{y=1,2,3,4}$ is assumed to be independent from $\rand{x}$. There are \emph{no} other restrictions on the joint distribution of $\rands{\varepsilon}$. This specification nests the logit specification estimated in the previous section. Hence, if the assumptions of multinomial logit are correct, then the results of parametric and semiparametric estimators should not differ much. 
\par
The estimates of $\beta_0$ and $\beta_1$ are $\hat\beta_0=-39.1$ (standard error$=43.8$) and $\hat\beta_1=-16.7\times 10^{-3}$ (standard error$=3.97\times 10^{-6}$).
\footnote{I use the tensor product of the $4$-th degree Chebyshev polynomials for $d$ and the $1$-st degree Chebyshev polynomials for every $z_{y}$.} 
Similar to the multinomial logit estimator, the sign of $\hat\beta_0$ is negative. The coefficient in front of the income variable is negative and significant at the $5$ percent significance level. However, the maximal value that $\hat{\beta}_1\rand{d}$ can take in the sample is substantially smaller than $\hat\beta_0$
($\max_i\left(\rand{d}^{(i)}\hat{\beta}_1 /\hat{\beta}_0\right)=0.055$, standard error$=0.062$). The latter indicates that, similarly to the fully parametric specification, income does not affect marginal disutility from price increase much. However, the estimate of $\beta_0$ is substantially lower than the one in the fully parametric case. This indicates that consumers may be less sensitive to price changes than one would think after estimating the logit-type model.  
\par
Interestingly, the difference between the estimates obtained using the fully parametric logit estimator $\bar{\beta}$ and my semiparametric estimator $\hat{\beta}$ is substantial (e.g., $\bar\beta_1/\hat\beta_1>10^3$). That is, the parametric estimator overestimates the magnitude of the agents sensitivity to relative price changes of margarine.  This suggests that the multinomial logit structure most likely fails to hold, emphasizing the importance of semiparametric estimation.\footnote{This empirical finding is in line with the simulation results, presented in Appendix~\ref{app: simulations}.}

\section{Conclusion}\label{sec:conclusion}
This paper shows that commonly used exclusion restrictions and richness assumptions about the distribution of some unobservables may lead to full nonparametric identification in discrete outcome models even when covariates are bounded. The proposed identification framework extends the results from a large literature that uses special covariates with full support to environments where such full-support covariates are not available. It also leads to an asymptotically normal estimator of the finite-dimensional parameters of the model.
  
\bibliography{references}
\appendix
\section{Proofs}\label{app: proofs}
I first establish identification of a more general model but without covariates $w$. This result will be used to prove the propositions from the main text. Assume that each instance of the environment is characterized by an endogenous outcome $\rand{y}$ from a known finite set $Y$, a vector of observed exogenous characteristics $\rand{x}\in X\subseteq\Real^{d_x}$, $d_x<\infty$, that can be partitioned into $x=(d,z)$, and a vector of unobserved indexes $\rand{s}\in S\subseteq\Real^{d_s}$.
\begin{assumption}[Data]\label{ass:app data}
There exists $Y^*\subseteq Y$ such that the analyst observes (can consistently estimate) 
$
\mu(y|x)=\Pr(\rand{y}=y|\rand{x}=x)
$
for all $x\in x$ and $y\in Y^*$.
\end{assumption}
\begin{assumption}\label{ass: app exclusion restriction general}
There exists $h_0: Y^*\times S\to[0,1]$, such that
    $
    \Pr(\rand{y}=y|\rand{x}=x,\rand{s}=s)=h_0(y,s),
    $
    for all $y\in Y^*$, $x\in X$, and $s\in S$.
\end{assumption}
Assumption~\ref{ass: app exclusion restriction general} is an exclusion restriction that requires $\rand{d}$ and $\rand{z}$ to affect distribution over outcomes in $Y^*$ only via the distribution of $\rand{s}$.
\begin{assumption}[Bounded completeness]\label{ass: app bounded completeness general}
There exists $X'\subseteq X$ such that the family of distributions $\left\{F_{\rand{s}|\rand{x}}(\cdot|x),x\in X'\right\}$ is boundedly complete.
\end{assumption}
\begin{proposition}\label{prop: app identification of h up to v}
Under Assumptions~\ref{ass:app data}-\ref{ass: app bounded completeness general}, $h_0$ is identified from $\mu$ up to $F_{\rand{s}|\rand{x}}$.
\end{proposition}
\begin{proof}
Fix some $y\in Y^*$. Under Assumption~\ref{ass: app exclusion restriction general}, I have the following integral equation
\[
\forall x\in X\::\:\mu(y|x)=\int_{S}h(y^*,s)dF_{\rand{s}|\rand{x}}(s|x).
\]
Suppose that there exists $h$ with $h(y^*,s)\neq h_0(y^*,s)$ for all $s$ in some nonzero-measure set $S'$ such that 
\[
\forall x\in X\::\:\mu(y|x)=\int_{S}h(y^*,s)dF_{\rand{s}|\rand{x}}(s|x)=\int_{S}\dop_0(y^*,s)dF_{\rand{s}|\rand{x}}(s|x).
\]
This implies that the nonzero function $h(y,\cdot)-h_0(y,\cdot)$ integrates to $0$ for all $x\in X'$. The latter contradicts to Assumption~\ref{ass: app bounded completeness general}. The fact that the choice of $y\in Y^*$ was arbitrary completes the proof. 
\end{proof}

\subsection{Nonparametric Identification}\label{sec: app nonparametric}
Given a collection of random variables $\{\rands{\xi}\}_{i=1,\dots,d}$, $d<\infty$, I say that $\rands{\xi}_i$ is redundant if there exists $j\neq i$ such that $\rands{\xi}_i=\rands{\xi}_j\:\as$. Nonredundant elements of $\{\rands{\xi}\}_{i=1,\dots,d}$ is the largest subset of $\{\rands{\xi}\}_{i=1,\dots,d}$ such that non of its elements are redundant.
\begin{assumption}\label{ass: app nonparametric errors}
\begin{enumerate}
\item The latent $\rand{s}=(\rand{s}_i)_{i=1,\dots,d_s}$ satisfies
\[
\rand{s}_i=\rand{z}_{i}[\beta_{0,i}+\beta_{1,i}\rand{d}_{i}+\rand{e}_i]\:\as
\]
where $\beta_{0,i}$ and $\beta_{1,i}$ are some unknown parameters such that $\beta_{1,i}\neq 0$ for all $i=1,\dots,d_{s}$;
\item Nonredundant elements of $\{\rand{e}_i\}_{i=1,\dots,d_s}$ are mean-zero and variance-one independent random variables that are independent of $\rand{x}$;
\item $h_0(y^*,\cdot)$ has bounded derivatives up to order $\kappa$ and $\partial^l_{s^l_i}h_0(y^*,\cdot)|_{s=0}\neq 0$ for all $l\leq\kappa$ and all $i=1,\dots,d_{s}$; 
\item The support of $\rand{x}$, which consists of nonredundant elements of $\{\rand{d}_i\}_{i=1,\dots,d_s}$ and all of $\{\rand{z}_i\}_{i=1,\dots,d_s}$, contains $x^*$ with an open neighborhood such that $z^*_{i}=0$ for all $i=1,\dots,d_{s}$;
\item The sign of either $\beta_{0,i}$ or $\beta_{1,i}$ is known for every $i=1,\dots,d_{s}$.
\end{enumerate}
\end{assumption}
\par
Let $\beta_0=\{\beta_{0,i}\}_{i=1}^{d_s}$ and $\beta_1=\{\beta_{1,i}\}_{i=1}^{d_s}$. 
\begin{proposition}\label{prop: app identification of beta2}
If Assumptions~\ref{ass:app data},~\ref{ass: app exclusion restriction general}, and~\ref{ass: app nonparametric errors} hold, then $\beta_{0}$, $\beta_1$, and $\Exp{\rand{e}^l_i}$, $i=1,\dots,d_s$, $0\leq l\leq\kappa$, are identified. 
\end{proposition}
\begin{proof}
Given a family $x = (x_k)_{k\in K}$ and a particular index value $k\in K$, let $x_{-k}$ denote $(x_j)_{j\in K\setminus\{k\}}$. Fix some $i\in\{1,2,\dots,d_s\}$ and set $z_{-i}$ to $0$. Take any $y^*\in Y^*$ from Assumption~\ref{ass: app exclusion restriction general}. To simplify notation, let $F_0:\Real\to\Real$ and $\eta:\Real^2\to\Real$ such that
$
F_0(t)=h_0(y^*,(0,\dots,t,\dots,0)),
$
where the only nonzero component in the second argument of $h_0$ is the $i$-th component, and $\eta(d_{i},z_{i})=\mu(y^*|x)$. Note that Assumption~\ref{ass: app nonparametric errors}(iii) together with the dominated convergence theorem imply that $F_0$ is has bounded derivatives up to order $\kappa$.
\par
Assumptions~\ref{ass: app exclusion restriction general}  implies that 
\begin{equation*}
\eta(d_{i},z_{i})=\int F_0((\beta_{0,i}+\beta_{1,i}d_{i}+e_i)z_{i})dF_{\rand{e}_i|\rand{x}}(e|x).
\end{equation*}
Next, since $\rand{e}_i$ and $\rand{x}$ are independent and $h_0(y^*,\cdot)$ is $\kappa$-times differentiable with bounded derivatives, the dominated convergence theorem implies that (I dropped the subscript $i$ from the notation)
\[
\partial^l_{d^l}\eta(d,z)=\beta_{1}^lz^l\int \partial^l_{t^l}F_0((\beta_{0}+\beta_{1}d+e)z)dF_{\rand{e}}(e)
\]
for any $l\leq\kappa$.
Hence, since derivatives of $h_0(y^*,\cdot)$ are bounded, applying the dominated convergence theorem again I get that
\[
\lim_{z\to 0}\dfrac{\partial^l_{d^l}\eta(d,z)}{z^l}=\beta_{1}^l\int \partial^l_{t^l}F_0(0)dF_{\rand{e}}(e)=\beta_{1}^l\partial^l_{t^l}F_0(0),
\]
and, thus, $\beta_{1}^l\partial^l_{t^l}F_0(0)$ is identified for any $l\leq\kappa$. Similarly note that, since $h_0(y^*,\cdot)$ has bounded derivatives,
\begin{equation}\label{eq:proof nonpara1}
\partial^l_{z^l}\eta(d,0)=\int \partial^l_{t^l}F_0(0)(\beta_{0}+\beta_{1}d+e)^ldF_{\rand{e}|\rand{x}}(e|x)
\end{equation}
for every $l\leq\kappa$. Hence, since $\Exp{\rand{e}}=0$ and $\beta_{1}\partial_{t}F_0(0)$ is identified,
$
\beta_{0}\partial_{t}F_0(0)=\partial_{z}\eta(d,0)-\beta_{1}\partial_{t}F_0(0)d
$
is also identified. Thus, we can identify $\beta_{0}/\beta_{1}$ and learn the sign of $\beta_{1}$ from Assumption~\ref{ass: app nonparametric errors}(v). For $l=2$, since $\Exp{\rand{e}}=0$ and $\Exp{\rand{e}^2}=1$, we also can derive that
\[
\partial^2_{z^2}\eta(d,0)=\int \partial^2_{t^2}F_0(0)(\beta_{0}+\beta_{1}d+e)^2dF_{\rand{e}|\rand{x}}(e|x)=\partial^2_{t^2}F_0(0)\left[(\beta_{0}+\beta_{1}d)^2+1\right].
\]
Hence,
$
\partial^2_{z^2}\eta(d,0)=\beta_{1}^2\partial^2_{t^2}F_0(0)\left[(\beta_{0}/\beta_{1}+d)^2+1/\beta_{1}^2\right].
$
As a result, since we identified $\beta_{0}/\beta_{1}$ and $\beta_{1}^2\partial^2_{t^2}F_0(0)$ in the previous steps,
\[
1/\beta_{1}^2=\dfrac{\partial^2_{z^2}\eta(d,0)}{\beta_{1}^2\partial^2_{t^2}F_0(0)}-(\beta_{0}/\beta_{1}+d)^2
\]
is identified. Since I already identified the sign of $\beta_{1}$ and $\beta_{0}/\beta_{1}$, I can identify $\beta_{0}$ and $\beta_{1}$. Moreover, I identify $\partial^l_{t^l}F_0(0)$ for all $l\leq\kappa$.
\par
To identify all moments of $\rand{e}$ up to order $\kappa$, I use Equation (\ref{eq:proof nonpara1}) to derive the following recursive equation
\[
\Exp{\rand{e}^l}=\frac{\partial^l_{z^l}\eta(d,0)}{\partial^l_{t^l}F_0(0)}-\sum_{k=1}^{l}{l \choose k}(\beta_{0}+d)^k\Exp{\rand{e}^{l-k}}.
\]
\par
Going back to the original notation, I identify $\beta_{0,i}$, $\beta_{1,i}$, and $\Exp{\rand{e}^l_i}$, $0\leq l\leq \kappa$. The conclusion of the proposition then follows from the fact that the choice of $i$ was arbitrary.
\end{proof}
Note that Proposition~\ref{prop: app identification of beta2} allows $\{\rand{z}_i\}_{i=1,\dots,d_v}$ and nonredundant elements of $\{\rand{e}_i\}_{i=1,\dots,d_v}$ and $\{\rand{d}_i\}_{i=1,\dots,d_v}$ to have different cardinality. If the cardinality of nonredundant elements of $\{\rand{e}_i\}_{i=1,\dots,d_v}$ and $\{\rand{d}_i\}_{i=1,\dots,d_v}$ is the same, then the assumption that $\{\rand{e}_i\}_{i=1,\dots,d_v}$ are independent can be relaxed. In this case, using a similar strategy, one can identify recursively $\Exp{\prod_{i\in I}\rand{e}^{\kappa_i}_i}$ for all possible $I\subseteq \{1,\dots,d_v\}$ and set of nonnegative integers $\{\kappa_i\}_{i\in I}$ such that $\sum_{i\in I}\kappa_i\leq \kappa$. For instance, if $d_v=2$, then for $F(v)=h(y^*,(v_1,v_2))$ I have that
\[
\eta(d,z)=\int_{\Real^2}F((\beta_{0,1}+\beta_{1,1}d_1+e_1)z_1,(\beta_{0,2}+\beta_{1,2}d_2+e_2)z_2)dF_{\rand{e}}(e).
\]
Thus, given that $\beta_0$ and $\beta_1$ are already identified, we can identify, $\partial^2_{t_1,t_2}F(0)$ since
\[
\lim_{\norm{z}\to 0}\dfrac{\partial^2_{d_1,d_2}\eta(d,z)}{z_1z_2}=\beta_{1,1}\beta_{1,2}\int \partial^2_{t_1,t_2}F(0)dF_{\rand{e}}(e)=\beta_{1,1}\beta_{1,2}\partial^2_{t_1,t_2}F(0).
\]
As a result, the partial derivative with respect to $z_1$ and $z_2$
\[
\partial^2_{z_1,z_2}\eta(d,0)=\int_{\Real^2}(\beta_{0,1}+\beta_{1,1}d_1+e_1)(\beta_{0,2}+\beta_{1,2}d_2+e_2)\partial^2_{t_1,t_2}F(0)dF_{\rand{e}}(e)
\]
identifies $\Exp{\rand{e}_1\rand{e}_2}$. Similarly, one can identify $\Exp{\rand{e}^{\kappa_1}_1\rand{e}^{\kappa_2}_2}$ for all possible positive integers $\{\kappa_i\}_{i=1,2}$ such that $\sum_{i=1,2}\kappa_i\leq \kappa$.
\subsection*{Normal Random Coefficient}
\begin{assumption}\label{ass: app normal errors}
\begin{enumerate}
\item The latent $\rand{s}=(\rand{s}_i)_{i=1,\dots,d_s}$ satisfies
\[
\rand{s}_i=\rand{z}_{i}[\beta_{0,i}+\beta_{1,i}\rand{d}_{i}+\rand{e}_i]\:\as,
\]
where $\beta_{0,i}$ and $\beta_{1,i}$ are some unknown parameters such that $\beta_{1,i}\neq 0$ for all $i=1,\dots,d_{s}$;
\item $\{\rand{e}_i\}_{i=1,\dots,d_s}$ are i.i.d. standard normal random variables that are independent of $\rand{x}$;
\item The support of $(\rand{d},\rand{z})$ contains an open ball;
\item The sign of either $\beta_{0,i}$ or $\beta_{1,i}$ is known for every $i=1,\dots,d_{s}$.
\end{enumerate}
\end{assumption}
The only support restriction is imposed on $\rand{d}$ and $\rand{z}$ (Assumption~\ref{ass: app normal errors}(iii)). Assumptions~\ref{ass: app normal errors}(i)-(iii) are sufficient for Assumption~\ref{ass: app bounded completeness general} since the family of normal distributions indexed by the mean is complete as long as the parameter space for the mean contains an open ball.
\par
Let $d_{-i}=(d_{k})_{k\neq i}$. For a fixed $y^*\in Y^*$, $d_{-i}$ and $z$, let $\eta:D_{i|d_{-i},z}\to[0,1]$ be such that for $x=((d_{i},d_{-i}),z)$,
$
\eta(d_{i})=\mu(\y^*|\x).
$
\begin{assumption}\label{ass: app support restriction}
For every $i=1,2,\dots,d_{s}$, there exists $y^*\in Y^*$ and $z_{i}\in Z_{i}\setminus\{0\}$ such that $\eta(\cdot)$ is neither an exponential nor an affine function.
\end{assumption}
\begin{proposition}\label{prop: app identification of beta}
Suppose that Assumptions~\ref{ass:app data},~\ref{ass: app exclusion restriction general},~\ref{ass: app normal errors}, and~\ref{ass: app support restriction} hold. Then $h_0$,  $\beta_{0}$, and $\beta_{1}$ are identified. 
\end{proposition}
\begin{proof}
Note that $\dop_0$ is identified up to $\beta_{0}$ and $\beta_{1}$ because of completeness of the family of normal distributions and Proposition~\ref{prop: app identification of h up to v}. Hence, I only need to show that $\beta_{0}$ and $\beta_{1}$ are identified.
Fix some $i\in\{1,2,\dots,d_s\}$, $z_{-i}$, and $d_{-i}$ in the support. Take $y^*$ from Assumption~\ref{ass: app support restriction}. To simplify notation, let $F_0:\Real\to\Real$ and $\eta:\Real^2\to\Real$ be functions such that
\[
F_0(s_i)=\int_{\Real^{d_s-1}}h_0(y^*,s)\prod_{k\neq i}\dfrac{\phi\left(s_k/z_{k}-\beta_{0,k}-\beta_{1,k}d_{k}\right)}{z_{k}}ds_{k},
\]
where $\phi(\cdot)$ is the standard normal p.d.f., and $\eta(d_{i},z_{i})=\mu(y^*|d,z)$. 
\par
Assumptions~\ref{ass: app exclusion restriction general} and~\ref{ass: app normal errors} imply that 
$
\eta(d_{i},z_{i})=\int_{\Real} F_0(s_i)\phi(s_i/z_{i}-\beta_{0,i}-\beta_{1,i}d_{i})ds_i/z_{i}.
$
After some rearrangements and dropping subscript $i$ from the notation, I get
\begin{equation}\label{eq:rc 1}
\tilde\eta(d,z)=\int_{\Real} F_0(s)\phi(s/z-\beta_{0}-\beta_{1}d)ds,
\end{equation}
where $\tilde\eta(d,z)=z\eta(d,z)$.
\par
Next, note that since $\partial^2_{x^2}\phi(x)=-\phi(x)-x\partial_x\phi(x)$ the following system of equations holds\footnote{I can differentiate under the integral sign since (i) $h_0$ being bounded implies that $F_0$ is bounded, (ii) all derivatives of the standard normal p.d.f. are bounded.}
\begin{align*}
\partial_{d}\tilde\eta(d,z)&=-\beta_{1}\int F_0(t)\partial_x\phi(t/z-\beta_{0}-\beta_{1}d)dt,\\
\partial^2_{d^2}\tilde\eta(d,z)&=\beta^2_1\int F_0(t)\partial^2_{x^2}\phi(t/z-\beta_{0}-\beta_{1}d)dt\\
&=-\beta^2_1\tilde\eta(d,z)-\beta_{1}(\beta_{0}+\beta_{1}d)\partial_{d}\tilde\eta(d,z)-\beta_{1}^2\int tF_0(t)\partial_x\phi(t/z-\beta_{0}-\beta_{1}d)dt/z.
\end{align*}
Moreover,
$\partial_{z}\tilde\eta(d,z)=-\int F_0(t)t\partial_x\phi(t/z-\beta_{0}-\beta_{1}d)dt/z^2$.
Hence,
\begin{align*}
\partial^2_{d^2}\tilde\eta(d,z)&=-\beta^2_1\tilde\eta(d,z)-\beta_{1}(\beta_{0}+\beta_{1}d)\partial_{d}\tilde\eta(d,z)+\beta_{1}^2z\partial_{z}\tilde\eta(d,z).
\end{align*}
Equivalently,
\begin{align*}
\dfrac{\beta_{0}}{\beta_{1}}=\dfrac{z\partial_{z}\tilde\eta(d,z)-\tilde\eta(d,z)}{\partial_{d}\tilde\eta(d,z)}-d -\dfrac{\partial^2_{d^2}\tilde\eta(d,z)}{\partial_{d}\tilde\eta(d,z)}\dfrac{1}{\beta_{1}^2}.    
\end{align*}
Replacing $\tilde\eta(d,z)$ by $z\eta(d,z)$, I get
\begin{align}\label{eq:proof main}
\dfrac{\beta_{0}}{\beta_{1}}=\dfrac{z\partial_{z}\eta(d,z)-d\partial_{d}\eta(d,z)}{\partial_{d}\eta(d,z)}-\dfrac{\partial^2_{d^2}\eta(d,z)}{\partial_{d}\eta(d,z)}\dfrac{1}{\beta_{1}^2}. \end{align}
Thus, $\beta_{0}/\beta_{1}$ is identified up to $\beta_{1}^2$.
Differentiating the last equation with respect to $d$ leads to the following equation
\begin{align}\label{eq: bet12}
\dfrac{1}{\beta_{1}^2}=\partial_{d}\left[\dfrac{z\partial_{z}\eta(d,z)-d\partial_{d}\eta(d,z)}{\partial_{d}\eta(d,z)}\right]/\partial_{d}\left[\dfrac{\partial^2_{d^2}\eta(d,z)}{\partial_{d}\eta(d,z)}\right].    
\end{align}
Hence, if
\begin{align}\label{eq:proof der1}
\partial_{d}\left[\dfrac{\partial^2_{d^2}\eta(d,z)}{\partial_{d}\eta(d,z)}\right]\neq0
\end{align}
for \emph{some} $d$ and $z$, then $\beta_{1}^2$ is identified. 
Suppose this is not the case. That is, for all $d$ and $z$
$
\partial_{d}\left[\dfrac{\partial^2_{d^2}\eta(d,z)}{\partial_{d}\eta(d,z)}\right]=0.
$
Equivalently,
$
\partial^2_{z^2}\left[\log(\partial_{d}\eta(d,z))\right]=0
$
for all $d$ and $z$.
The latter would imply that either 
$
\eta(d,z)=K_1(z)e^{K_3(z)d}+K_2(z)
$
or
$
\eta(d,z)=K_1(z)d+K_2(z)
$
for some functions $K_k(\cdot)$, $k=1,2,3$. Since it is assumed that $\eta(\cdot,z)$ is neither an exponential nor an affine function on some open set, I can conclude that for some $d$ and $z$ Equation (\ref{eq:proof der1}) is satisfied. Thus, $\beta_{1}^2$ is identified (hence, $\abs{\beta_{1}}$ is also identified). Hence, I identify $\beta_{0}/\beta_{1}$. If $\beta_{0}/\beta_{1}=0$, then the sign of $\beta_{1}$ is identified from Assumption~\ref{ass: app normal errors}(iv). If $\beta_{0}/\beta_{1}\neq 0$, then the sign of either $\beta_{1}$ or $\beta_{0}$ is identified from Assumption~\ref{ass: app normal errors}(iv). Knowing the sign of, say, $\beta_{0}$ and $\beta_{0}/\beta_{1}$ identifies $\beta_{1}$ and $\beta_{0}$. Going back to the original notation I identify $\beta_{1,i}$ and $\beta_{0,i}$. The conclusion of the proposition then follows from the fact that the choice of $i$ was arbitrary.  
\par
Note that for identification of $\beta_{1}$ and $\beta_{0}$, I do not need to exclude all exponential functions of $d$, since instead of differentiating Equation (\ref{eq:proof main}) with respect to $d$, I can differentiate it with respect to $z$. For the identification result to hold it suffices to exclude functions of the form
$\eta(d,z)=K_1(z)e^{K_2d}+K_3(z)$ or $\eta(d,z)=K_1(z)d+K_3$,
\end{proof}

\subsection{Proof of Propositions~\ref{prop:nonparameric identification1},~\ref{prop:nonparameric identification2}, and~\ref{prop: normal mult choice}}
In the previous section, I stated and proved two general identification results (Propositions~\ref{prop: app identification of beta2} and~\ref{prop: app identification of beta}). Next I will apply these results to a multinomial choice model studied in the main text of the paper.
\par
Fix some arbitrary $w\in W$. To prove Propositions~\ref{prop:nonparameric identification1} and~\ref{prop: normal mult choice}(i), I use Propositions~\ref{prop: app identification of beta2} and~\ref{prop: app identification of beta}. Both propositions require Assumptions~\ref{ass:app data} and~\ref{ass: app exclusion restriction general}. Assumptions~\ref{ass:app data} is implied by Assumption~\ref{ass:observables} for $Y^*=\{0\}$. Assumption~\ref{ass: app exclusion restriction general} is satisfied in Propositions~\ref{prop:nonparameric identification1} with $h(0,s)=F_{\rands{\varepsilon}|\rand{w}}(0,\dots,s,\dots,0)$, where the the only nonzero component corresponds to $\bar{y}$ from Assumption~\ref{ass:random coeff nonparametric}(ii). To show validity of Assumption~\ref{ass: app exclusion restriction general} in Proposition~\ref{prop: app identification of beta}, note that under Assumption~\ref{ass:random coeff nonparametric}.(iii) or Assumption~\ref{ass:normal error random coef}.(ii) there exists $z^*$ and $\{\lambda_{y}\}_{y=1}^{J}$ with some open neighbourhood such that $z^*_{y'}=\lambda_{y'}z^*_{1}$ for all $y'\in Y$ with $\min_{y'}\lambda_{y'}>0$. 
Note that since $\rand{e}$ and $\rand{z}$ are independent conditional on $\rand{w}$, I have that for $x^*=(d^*,z^{*},w)$
\begin{equation}\label{eq: sign1}
\Scale[0.9]{\mu(0|x^*)=\int_{\Real}F_{\rands{\varepsilon}|\rand w}(-z^*_{1}(\beta_0(w)+\beta_{1}(w)d^*+e),\dots,-\lambda_{J}z^*_{1}(\beta_0(w)+\beta_{1}(w)d^*+e)|w)dF_{\rand{e}|\rand{w}}(e|w).}
\end{equation}
Hence, Assumptions~\ref{ass: app exclusion restriction general} is satisfied for $h(0,s)=F_{\rands{\varepsilon}|\rand{w}}(s,\lambda_2s,\cdot,\lambda_{J}s|w)$. 
The rest of assumptions follow from Assumption~\ref{ass:random coeff nonparametric} or Assumption~\ref{ass:normal error random coef}, except for identification of the sign of $\beta_0$ or $\beta_1$. However, I can identify the sign of $\beta_{1}(w)$ from Equation (\ref{eq: sign1}) since $F_{\rands{\varepsilon}|\rand{w}}(\cdot|w)$ is weakly monotone. As a result, I can identify $\beta_0(w)$, $\beta_1(w)$, and $\Exp{\rand{e}^l|\rand{w}=w}$, $0\leq l\leq\kappa$ (if $\rand{e}$ is standard normal then we already know its distribution). The fact that the choice of $w$ was arbitrary completes the proof.
\par
To prove Propositions~\ref{prop:nonparameric identification2} and~\ref{prop: normal mult choice}(ii), note that since $\beta_0$, $\beta_{1}$, and $F_{\rand{e}|\rand{x}}$ are identified (either from its moments or because it is standard normal), I know the distribution of $\rand{v}=\beta_0(\rand{w})+\beta_{1}(\rand{w})\rand{d}+\rand{e}$. Moreover, $F_{\rand{v}|\rand{x}}$ constitutes a boundedly complete family either by the assumption in Proposition~\ref{prop:nonparameric identification2} or by normality of $\rand{e}$ and continuity of $\rand{d}$ in an open ball \citep{brown86}. Hence, since
\begin{align*}
\Pr(\rand{y}=0|\rand{x}=x)&=\int_{\Real}F_{\rands{\varepsilon}|\rand{w}}(-z_{1}v,\dots,-z_{J}v|w)dF_{\rand{e}|\rand{w}}(v-\beta_0(w)-\beta_{1}(w)d|w)=\\
&=\int_{\Real}\tilde{g}(z,w,v)dF_{\rand{e}|\rand{w}}(v-\beta_0(w)-\beta_{1}(w)d|w)
\end{align*}
and Assumptions~\ref{ass: app exclusion restriction general} is satisfied, I can identify
$
\tilde{g}(z,w,v)=F_{\rands{\varepsilon}|\rand{w}}(-z_{2}v,\dots,-z_{J}v|w)
$
for all $z,w,v$ by Proposition~\ref{prop: app identification of h up to v}. Note that since $v$ can take any value in 
\[
V_{w}=\{v\::\:v=e+\beta_1(w)d+\beta_0(w), e\in E_{w},d\in D_{w}\}
\]
for any direction $-z/\norm{z}$ in the support of $\rand{z}$ conditional on $\rand{w}=w$, I can recover $F_{\rands{\varepsilon}|\rand{w}}(g|w)$ for any $g$ such that $g=-zv/\norm{z}$ for some $v$. That is, I identify $F_{\rands{\varepsilon}|\rand{w}}(\cdot|w)$ over the set $R_w$.

\subsection{Proof of Proposition~\ref{prop: estimator}}\label{app: proof of est}
To simplify the notation, I will focus on the binary choice case.
\par
\noindent\emph{Step 1.} In this step I make several observations about $p_0$ and its derivatives. By definition $0\leq h_0(v)\leq1$ for all $v$ and
\[
p_0(x)=\int_{\Real}h_0((\beta_0+\beta_1 d+e)z_{1})\phi(e)de=\int_{\Real}h_0(v)\phi(v/z_{1}-\beta_1 d-\beta_0)dv/z_{1}.
\]
Hence, $p_0$ is continuously differentiable of any order. Moreover, $p_0(x)=0$ if and only if $h(v)=0$ for all $v$. The latter means that probability of picking the outside option conditional on $\rand{x}=x$ and $\rand{e}=e$ equals to $0$ for all $e$. Since $\rands{\varepsilon}_1$ is independent of $\rand{x}$ and $\rand{e}$, I have that
$
\rands{\varepsilon}_1\geq -z_{1}(\beta_0+\beta_1 d+e)
$
with probability $1$ for all $e$, which is not possible since $\rand{e}$ has full support. Thus, $p_0(x)>0$ for all $x$. Similarly, one can show that $p_0(x)<1$ for all $x$.
\par
Next consider $p_1(x)=\partial_{d}p_0(x)$. Since $\partial_t\phi(t)=-t\phi(t)$,
\begin{align*}
\abs{p_1(x)}&=\abs{\beta_1\int_{\Real}h_0(v)(v/z_{1}-\beta_1 d-\beta_0)\phi(v/z_{1}-\beta_1 d-\beta_0)dv/z_{1}}=\abs{\beta_1\int_{\Real}h_0((\beta_0+\beta_1 d+e)z_{1})e\phi(e)de}.
\end{align*}
Hence, since $0\leq h_0(v)\leq 1$ for all $v$, I get that for some $C_1<\infty$,
$
\sup_{x}\abs{p_1(x)}\leq\beta_1\int_{\Real}\abs{e}\phi(e)de\leq C_1.
$
Similarly, note that $p_2(x)=z_{1}\partial_{z_{1}}p_0(x)$ and by the triangular inequality
\begin{align*}
\abs{p_2(x)}\leq\abs{p_0(x)}+\abs{\int_{\Real}h_0((\beta_0+\beta_1 d+e)z_{1})e\phi(e)(\beta_0+\beta_1 d+e)de}.
\end{align*}
Hence, given bounded support of $x$, I can conclude that $\sup_x\abs{p_2(x)}$ is also finite. Repeating the above steps, one can show that all higher order partial derivatives of $p_0$ are bounded.
\par
\noindent\emph{Step 2.}
Note that in the proof of Proposition~\ref{prop: normal mult choice} we used derivatives of $\eta(d,z_{1})$ to identify $\beta$s. In particular, we can take
$
\eta(d^{**},z^{**}_{1})=\mu(0|x^{**}),
$
where $x^{**}=(d^{**},(\lambda_{y}z^{**}_{1}))_y,w)$ and $\lambda_y=z^{**}_{2,y}/z^{**}_{2,1}$. As a result,  
$
\partial_{z_{1}}\eta(d^{**},z^{**}_{1})=\sum_y\lambda_y\partial_{z_{y}}\mu(0|x^{**}).
$
Since $\lambda_y=z^{**}_{2,y}/z^{**}_{2,1}$, I get that
$
z^{**}_{1}\partial_{z_{1}}\eta(d^{**},z^{**}_{1})=\sum_y z^{**}_{y}\partial_{z_{y}}\Pr(\rand{y}=0|\rand{x}=x^{**}).
$
Hence, if Assumption~\ref{ass:normal error random coef}(iii) is satisfied not just for one $(d^{**},z^{**})$ but for all, then for all $x$
\begin{align}\label{eq:beta for estimation}
    \nonumber\beta^2_{1}&=\dfrac{\partial^3_{d^3}p_0(x)\partial_{d}p_0(x)-[\partial_{d}p_0(x)]^2}{\sum_{y}z_{y}\partial^2_{d,z_{y}}p_0(x)\partial_{d}p_0(x)-\sum_{y}z_{y}\partial_{z_{y}}p_0(x)\partial^2_{d^2}p_0(x)-[\partial_{d}p_0(x)]^2},\\
    \beta_{0}&=\dfrac{\sum_{y}z_{y}\partial_{z_{y}}p_0(x)-d\partial_{d}p_0(x)}{\partial_{d}p_0(x)}\beta_{1}-\dfrac{\partial^2_{d^2}p_0(x)}{\partial_{d}p_0(x)}\dfrac{1}{\beta_{1}}.    
\end{align}

\noindent\emph{Step 3.} Combining the bounds for the derivatives from Step 1, the uniform weak law of large numbers, and consistency of $\hat{p}_0$, I can deduce that 
\begin{align*}
    &\dfrac{1}{n}\sum_{i=1}^n\hat{p}_{111}\left(\rand{x}^{(i)}\right)\hat{p}_1\left(\rand{x}^{(i)}\right)-\hat{p}_{11}\left(\rand{x}^{(i)}\right)^2\to_p\Exp{p_{111}(\rand{x})p_{1}(\rand{x})-p_{11}(\rand{x})^2},\\
    &\dfrac{1}{n}\sum_{i=1}^n\hat{p}_{12}\left(\rand{x}^{(i)}\right)\hat{p}_1\left(\rand{x}^{(i)}\right)-\hat{p}_2\left(\rand{x}^{(i)}\right)\hat{p}_{11}\left(\rand{x}^{(i)}\right)-\hat{p}_1\left(\rand{x}^{(i)}\right)^2\to_p\Exp{p_{12}(\rand{x})p_{1}(\rand{x})-p_{2}(\rand{x})p_{11}(\rand{x})-p_{1}(\rand{x})^2},\\
    &\dfrac{1}{n}\sum_{i=1}^n\hat{p}_{2}\left(\rand{x}^{(i)}\right)-\rand{d}^{(i)}\hat{p}_{1}\left(\rand{x}^{(i)}\right)\to_p\Exp{p_{2}(\rand{x})-\rand{d}p_{1}(\rand{x})},\\
    &\dfrac{1}{n}\sum_{i=1}^n\hat{p}_{11}\left(\rand{x}^{(i)}\right)\to_p\Exp{p_{11}(\rand{x})},\quad\quad
    \dfrac{1}{n}\sum_{i=1}^n\hat{p}_{1}\left(\rand{x}^{(i)}\right)\to_p\Exp{p_{1}(\rand{x})}.
\end{align*}
Thus, Equation (\ref{eq:beta for estimation}) and the continuous mapping theorem imply that $\hat\beta\to_p\beta$.
\par
\noindent\emph{Step 4.} Consider 
\[
\mathcal{G}_n=\dfrac{1}{n}\sum_{i=1}^n\left(\begin{array}{c}
    \hat{p}_{111}\left(\rand{x}^{(i)}\right)\hat{p}_1\left(\rand{x}^{(i)}\right)-\hat{p}_{11}\left(\rand{x}^{(i)}\right)^2\\
     \beta_1^2\left[\hat{p}_{2}\left(\rand{x}^{(i)}\right)-\rand{d}^{(i)}\hat{p}_{1}\left(\rand{x}^{(i)}\right)\right]-\hat{p}_{11}\left(\rand{x}^{(i)}\right)
\end{array}\right).
\]
To prove asymptotic normality of $\mathcal{G}_n$, I will use Theorem~6 in \citet{newey1997convergence}. The data is assumed to be i.i.d., the outcome variable is finite and $p_0$ is bounded and bounded away from $0$. Hence, Assumptions~1 and~4 from \citet{newey1997convergence} are satisfied. Assumption~8 in  \citet{newey1997convergence} is assumed. Assumption~9 in \citet{newey1997convergence} follows from Step 1. Finally, consider $a(p_0)=(a_1(p_0),a_0(p_0))$ with
\begin{align*}
    a_1(p_0)=\Exp{p_{111}(\rand{x})p_{1}(\rand{x})-p_{11}(\rand{x})^2},\quad\quad
    a_2(p_0)=\Exp{\beta_1^2[p_{2}(\rand{x})-\rand{d}p_{1}(\rand{x})]-p_{11}(\rand{x})}.
\end{align*}
The directional derivative of $a$ at $p_0$ in direction $g_0$ is then $D(g_0)=(D_1(g_0),D_2(g_0))$ with
\begin{align*}
    D_1(g_0)=\Exp{p_{111}(\rand{x})g_{1}(\rand{x})+g_{111}(\rand{x})p_{1}(\rand{x})-2p_{11}(\rand{x})g_{11}(\rand{x})},\quad
    D_2(g_0)=\Exp{\beta_1^2[g_{2}(\rand{x})-\rand{d}g_{1}(\rand{x})]-g_{11}(\rand{x})}.
\end{align*}
Applying integration by parts several times and using the fact that $f_{\rand{x}}$ and its partial derivatives vanish at the boundary of the support of $\rand{x}$ (Assumption~\ref{ass: regularity}(iii)), I get
\begin{align*}
    \Exp{p_{111}(\rand{x})g_{1}(\rand{x})}&=-\Exp{\partial_{z_1}[p_{111}(\rand{x})f_{\rand{x}}(\rand{x})]g_0(\rand{x})/f_{\rand{x}}(\rand{x})},\\
    \Exp{p_{1}(\rand{x})g_{111}(\rand{x})}&=-\Exp{\partial^3_{z^3_1}[p_{1}(\rand{x})f_{\rand{x}}(\rand{x})]g_0(\rand{x})/f_{\rand{x}}(\rand{x})},\\
    \Exp{p_{11}(\rand{x})g_{11}(\rand{x})}&=\Exp{\partial^2_{z^2_1}[p_{11}(\rand{x})f_{\rand{x}}(\rand{x})]g_0(\rand{x})/f_{\rand{x}}(\rand{x})},\\
    \Exp{\rand{d}g_{1}(\rand{x})}&=-\Exp{(f_{\rand{x}}(\rand{x})+\rand{d}\partial_{z_1}f_{\rand{x}}(\rand{x}))g_0(\rand{x})/f_{\rand{x}}(\rand{x})},\\
    \Exp{g_{11}(\rand{x})}&=\Exp{\partial^2_{z^2_1}f_{\rand{x}}(\rand{x})g_0(\rand{x})/f_{\rand{x}}(\rand{x})},\\
    \Exp{g_{2}(\rand{x})}&=-\Exp{(f_{\rand{x}}(\rand{x})+\rand{z}_{1}\partial_{z_2}f_{\rand{x}}(\rand{x}))g_0(\rand{x})/f_{\rand{x}}(\rand{x})}.
\end{align*}
As a result,
\begin{align*}
    D_1(g_0)&=-\Exp{\{4p_{1111}(\rand{x})f_{\rand{x}}(\rand{x})+8p_{111}(\rand{x})\partial_{z_1}f_{\rand{x}}(\rand{x})+5p_{11}(\rand{x})\partial^2_{z^2_1}f_{\rand{x}}(\rand{x})+p_{1}(\rand{x})\partial^3_{z^3_1}f_{\rand{x}}(\rand{x})\}g_0(\rand{x})/f_{\rand{x}}(\rand{x})},\\
    D_2(g_0)&=\Exp{\{\beta_1^2[\rand{d}\partial_{d}f_{\rand{x}}(\rand{x})-\rand{z}_{1}\partial_{z_2}f_{\rand{x}}(\rand{x})]-\partial^2_{d^2}f_{\rand{x}}(\rand{x})\}g_0(\rand{x})/f_{\rand{x}}(\rand{x})}.
\end{align*}
Hence,
$
D(g_0)=\Exp{\bar{v}(\rand{x})g_0(\rand{x})}.
$
Moreover, $\bar{v}$ is continuously differentiable and $\Exp{\bar{v}(\rand{x})\bar{v}(\rand{x})\tr}$ is finite and nonsigular (Assumption~\ref{ass: regularity}(iv)). Hence, Assumption~7 in \citet{newey1997convergence} is also satisfied, thus, by Theorem~6 in \citet{newey1997convergence},
$
\sqrt{n}\left(\mathcal{G}_n-\mathcal{G}\right)\to_d N(0,\tilde V),
$
where 
\[
\mathcal{G}=\Exp{\begin{array}{c}
    p_{111}(\rand{x})p_1(\rand{x})-p_{11}(\rand{x})^2\\
     \beta_1^2\left[p_{2}(\rand{x})-\rand{d}p_{1}(\rand{x})\right]-p_{11}(\rand{x})
\end{array}}
\]
and 
$
\tilde V=\Exp{\bar{v}(\rand{x})\bar{v}(\rand{x})\tr p_0(\rand{x})(1-p_0(\rand{x}))}.
$
Moreover, I can construct a consistent estimator of $\tilde{V}$ using Theorem~6 in \citet{newey1997convergence}. In particular, let $\hat{a}(\hat{p}_0)$ be a sample counterpart of $a(p_0)$ and
\begin{align*}
    \hat{\gamma}&=\left(\Psi\tr\Psi\right)^{-}\sum_{i=1}^n\psi^K\left(\rand{x}^{(i)}\right)\Char{\rand{y}^{(i)}=0},\quad\quad
    \hat{A}=\partial_{\gamma}\hat{a}(\psi^K(z)\tr\hat\gamma),\\
    \hat{Q}&=\Psi\tr\Psi/n,\quad\quad
    \hat{\Sigma}=\sum_{i=1}^n\psi^K\left(\rand{x}^{(i)}\right)\psi^K\left(\rand{x}^{(i)}\right)\tr\left[\Char{\rand{y}^{(i)}=0}-\hat{p}_0\left(\rand{x}^{(i)}\right)\right]^2/n.
\end{align*}
Then 
$
\hat{\tilde{V}}=\hat{A}\tr\hat{Q}^{-}\hat{\Sigma}\hat{Q}^{-}\hat{A}\to_p \tilde V.
$
\par
\noindent\emph{Step 5.} Combining Step 2 with the continuous mapping theorem, Slutsky's theorem, and the Delta method, implies that
\[\Scale[0.8]{
\sqrt{n}\left(\hat\beta-\beta\right)\to_{d}\left(\begin{array}{cc}
    2\beta_1 &  0\\
    0 & 1
\end{array}\right)\left(\begin{array}{cc}
    {\Exp{p_{12}(\rand{x})p_{1}(\rand{x})-p_{2}(\rand{x})p_{11}(\rand{x})-p_{1}(\rand{x})^2}} &  0\\
    0 & {\beta_1\Exp{p_{1}(\rand{x})}}
\end{array}\right)^{-1}N\left(0,\tilde V\right).}
\]
\par
\noindent\emph{Step 5.} Consistency of 
$
\hat{V}=\hat{G}\hat{\tilde{V}}\hat{G}\tr,
$
where
\[\Scale[0.8]{
\hat{G}=\left(\begin{array}{cc}
    2\hat\beta_1 &  0\\
    0 & 1
\end{array}\right)\left(\begin{array}{cc}
    {n^{-1}\sum_{i=1}^n{\hat{p}_{12}\left(\rand{x}^{(i)}\right)\hat{p}_{1}\left(\rand{x}^{(i)}\right)-\hat{p}_{2}\left(\rand{x}^{(i)}\right)\hat{p}_{11}\left(\rand{x}^{(i)}\right)-\hat{p}_{1}\left(\rand{x}^{(i)}\right)^2}} &  0\\
    0 & {n^{-1}\hat{\beta}_1\sum_{i=1}^n{\hat{p}_{1}\left(\rand{x}^{(i)}\right)}}
\end{array}\right)^{-1},}
\]

follows from consistency of $\hat\beta$, $\hat{\tilde{V}}$, Step 3, and the continuous mapping theorem.
\section{Additional Simulations}\label{app: simulations}
Table~\ref{table: mad my} contains results for the mean absolute deviation (MAD) of my estimator of $\beta_1$. Similar to the bias, the MAD decreases with $n$ and is of the similar magnitude across DGPs.
\begin{table}[h]
\centering
\begin{threeparttable}
\centering
\caption{Mean Absolute Deviation}\label{table: mad my}
\begin{tabular}{c|ccccccc}
\hline
\hline
Sample Size/DGP & DGP-$0$ & DGP-$1$ & DGP-$2$ & DGP-$3$ & DGP-$4$ & DGP-$5$ & DGP-L\\ 
\hline
1000            & 1.11    & 1.13    & 1.20    & 1.48    & 1.53    & 1.52    & 1.15 \\
5000            & 0.48    & 0.65    & 0.95    & 1.09    & 1.28    & 1.25    & 0.71 \\
10000           & 0.38    & 0.42    & 0.67    & 0.90    & 1.13    & 1.21    & 0.52 \\
\hline
\end{tabular}
\vspace{1ex}
\end{threeparttable}
\end{table}
\par
Next, I estimated $\beta_1$ using two maximum-likelihood estimators. The first one (Probit) is based on the assumption that $\rands{\varepsilon}$ is standard normal. The second one (Logit) is assumes that $\rands{\varepsilon}$ has a logistic distribution. The Probit estimator is correctly specified under DGP-$0$ and is misspecified for all other DGPs. The Logit estimator is misspecified for all DGPs except DGP-L. The results for the bias and the MAD for both estimators for $n=1000$ are presented in Tables~\ref{table: probit} and~\ref{table: logit}.
\begin{table}[h]
\centering
\begin{threeparttable}
\centering
\caption{Bias and Mean Absolute Deviation of the Probit estimator}\label{table: probit}
\begin{tabular}{c|ccccccc}
\hline
\hline
Metric/DGP & DGP-$0$ & DGP-$1$ & DGP-$2$ & DGP-$3$ & DGP-$4$ & DGP-$5$ & DGP-L\\ 
\hline
Bias       & 0.05    & 26.0    & 46.25    & 183.18    & 716.74    & 2197.74    & 25.05 \\
MAD        & 0.14    & 26.1    & 46.35    & 183.28    & 716.82    & 2197.81    & 25.19 \\
\hline
\end{tabular}
\vspace{1ex}
\end{threeparttable}
\end{table}

\begin{table}[h]
\centering
\begin{threeparttable}
\centering
\caption{Bias and Mean Absolute Deviation of the Logit estimator}\label{table: logit}
\begin{tabular}{c|ccccccc}
\hline
\hline
Metric/DGP & DGP-$0$ & DGP-$1$ & DGP-$2$ & DGP-$3$ & DGP-$4$ & DGP-$5$ & DGP-L\\ 
\hline
Bias       & 0.06    & 0.25    & 0.66    & 2.76    & 7.76    & 16.96    & 0.47 \\
MAD        & 0.15    & 0.34    & 0.77    & 2.85    & 7.84    & 17.01    & 0.59 \\
\hline
\end{tabular}
\vspace{1ex}
\end{threeparttable}
\end{table}
Overall, the Logit estimator outperforms the Probit estimator for all DGPs except DGP-$0$.  As expected, since for DGP-$0$ and DGP-L the Probit and the Logit estimators are correctly specified, respectively, the bias and the MAD are small and both estimators perform better than my estimator (see also Table~\ref{table: bias my}). However, for the rest of DGPs, these estimators perform very poorly. For instance, the bias of the Logit estimator is about 11 times bigger that the bias of my estimator for DGP-$5$. 
\end{document}